\documentclass[sn-mathphys]{sn-jnl}

\jyear{2021}%

\theoremstyle{thmstyleone}%
\theoremstyle{thmstyletwo}%
\newtheorem{Remark}{Remark}%
\newtheorem{Lemma}{Lemma}%
\newtheorem{Definition}{Definition}%
\newtheorem{Corollary}{Corollary}%

\usepackage{changepage}
\usepackage{fullwidth}

\theoremstyle{thmstylethree}%

\begin{document}

\title[    ]{Comparison of information criteria for detection of useful signal in noisy environment}


\author[1]{\fnm{Leonid} \sur{Berlin}}\email{berlin.lm@phystech.edu}
\equalcont{These authors contributed equally to this work.}

\author[1]{\fnm{Andrey} \sur{Galyaev}}\email{galaev@ipu.ru}
\equalcont{These authors contributed equally to this work.}

\author*[1]{\fnm{Pavel} \sur{Lysenko}}\email{pashlys@yandex.ru}
\equalcont{These authors contributed equally to this work.}

\affil*[1]{\orgdiv{Laboratory 38}, \orgname{Institute of Control Sciences of RAS},\\ \orgaddress{\street{} \city{Moscow}, \postcode{}\state{}\country{Russia}}}


\abstract{This paper considers the problem of appearance indication of useful acoustic signal in the signal/noise mixture. Various information characteristics (information entropy, Jensen-Shannon divergence, spectral information divergence and statistical complexity) are investigated in the context of solving this problem. Both time and frequency domain are studied for information entropy calculation. The effectiveness of statistical complexity is shown in comparison with other information metrics for different levels of added white noise. In addition analytical formulas for complexity and disequilibrium are obtained using entropy variation in the cases of one- and two-dimensional spectral distributions. The effectiveness of the proposed approach is shown for different types of acoustic signals and noise level especially when additional noise characteristics estimation is impossible. 
}




\maketitle

\section{Introduction}

Since Shannon in {\cite{shannon} introduced the concepts of information and information entropy, they have attracted a vast attention of scientists, as evidenced by a large number of articles devoted to the development of information theory in various theoretical and practical aspects. Many rather different information criteria, metrics and methods for their calculation, one way or another based on the concepts of Shannon entropy, have been proposed and investigated \cite{ent_uni}.
These metrics can be used quite successfully in signal processing, which eventually led to the emergence of a separate section of this scientific area, called entropic signal analysis \cite{gray_entropy_2011}.


For signals described by time series, the information entropy can be calculated both on the basis of signal representation in the time domain \cite{SampleEntropy} and on the basis of its representation in the frequency domain \cite{Shen1998RobustEE}, i.e. using the signal spectrum. The convenience of the second approach follows from the fact that white noise, which is usually used to model the background noise in these problem statements, has a uniform frequency distribution. That allows to simplify its mathematical description and solve the problem of separation of useful signal more effectively. 


Decision theory considers change-point detection problems, which are closely related to the problems discussed above: most often in such problems it is required to determine the moment of change in the parameters of a random process registered in discrete time. In \cite{book_Shiryaev, Moriarty} many probabilistic-statistical methods of solving such problems are considered. Also one can not ignore the so-called Anomaly Detection Problems, where it is required to detect the anomaly of time series \cite{mehrotra_anomaly_2017, e22080845, e17042367}, that is the moment when the behavior of the system begins to qualitatively differ from normal for various reasons, in particular due to external unwanted interference. Electrocardiogram (ECG) is one example of such time series, which is analyzed in a great number of articles, for example in \cite{TakuyaHorie}. The presence of an anomaly in this case can indicate health problems and its detection at an early stage may save the life of the patient. 


Of particular interest is the processing of acoustic signals, which can be useful, for example, in Voice Activity Detection (VAD) problems \cite{vad}, relevant to voice assistants. The task is usually to separate the speech segments from the background of environmental noise. The articles \cite{Shen1998RobustEE, RobustEndpoint, Weaver} present a method for endpoint detection, i.e. determining the limits of a speech signal in a mixture of this signal and background noise, based on the calculation of the spectral entropy. The general idea of methods based on information criteria is that their values experience a sharp jump when a useful signal appears in the noise.


In a series of articles \cite{lopez-ruiz_shannon_2005, catalan_features_2002, calbet_tendency_2001, chaos, LAMBERTI2004119, rosso} the researchers introduced the concept of a statistical measure of signal complexity, which they called statistical complexity. In \cite{features, plane} statistical complexity and information entropy are used to classify various underwater objects of animate and inanimate nature from the recorded sound. In present article, we propose to use this measure to indicate the appearance of an useful acoustic signal in a highly noisy mixture. It should be noted that the positive side of the proposed method is that it does not require any a priori knowledge about the signal to be detected. However, if the a priori information such as the approximate frequency range of the signal is known, its detection will be even more accurate.




The structure of the paper is as follows. Section \ref{Criter} provides a brief theoretical summary of the information criteria used in various known signal detection methods. In Section \ref{EntVariation} entropy variation is investigated and statistical complexity is introduced for one- and two-dimensional distributions justifying the proposed approach. Section \ref{Modelling} is devoted to demonstrating a variety of examples, comparison of different information criteria and discussion of these results which allows us to make an educated choice of a suitable rule for detection of the signal in a noisy mixture. Conclusion \ref{Conclusion} summarizes the conducted research and provides direction for future work.

\section{Information criteria}\label{Criter}
\subsection{Information entropy and other information criteria}
In information theory, the entropy of a random variable is the average level of ''surprise'' or ''uncertainty'' inherent in the possible outcomes of the variable. For a discrete random variable $X$ that takes values in the alphabet ${\mathcal X}$ and has distribution density $p:{\mathcal X} \to [0,1]$, the entropy, according to Shannon \cite{shannon}, is defined as
\begin{equation}\label{basic_entropy}
{\displaystyle \mathrm {H} (p):=-\sum _{x\in {\mathcal {X}}}p(x)\log_2 p(x)},
\end{equation}
where $\displaystyle \Sigma$ denotes the sum across all possible values of the variable. It follows from the formula \eqref{basic_entropy} that entropy reaches its maximum value when all the states of the system are equally probable.

There are several definitions of information divergences, i.e., statistical distances between two distributions. The Kullback-Leibler divergence (or mutual entropy) between two discrete probability distributions $p(x)$ and $q(x)$ on an event set $\mathcal{X}$ is defined as
\begin{equation}
    D_{KL}(p||q) = \sum_{x\in {\mathcal {X}}} p(x) \log_2\frac{p(x)}{q(x)}.
\end{equation}
This measure is a statistical distance and distinguishes statistical processes by indicating how much $p(x)$ differs from $q(x)$ by the maximum likelihood hypothesis test when the actual data obey the distribution $p(x)$.
It is easy to see that
\begin{equation}
D_{KL}(p||q) = H(p, q) - H(p), 
\end{equation}
where $H(p,q)$ is a cross-entropy between $p$ and $q$:
\begin{equation}
    {\displaystyle H(p,q)=\mathbb {E}_{p}[-\log_2 q]},
\end{equation}
where ${\displaystyle \mathbb {E}_{p}[\cdot ]}$ is an operator of the mathematical expectation relative to the distribution $p$.

The symmetrized Kullback-Leibler distance \cite{vad} is often used in studies:
\begin{equation}\label{symmer_Kullback}
   \rho(p||q) = D_{KL}(p||q) + D_{KL}(q||p).
\end{equation}

However, the Jensen-Shannon divergence, which symmetrizes the Kullback-Leibler divergence and is often a more convenient information measure for practical applications, is used more often:
\begin{equation}\label{J-S}
JSD(p||q) = \frac{D_{KL}(p||m) + D_{KL}(q||m)}{2}, \quad m = \frac{p+q}{2}.
\end{equation}

It is symmetric and always has a finite value. The square root of the Jensen-Shannon divergence is a metric often called the Jensen-Shannon distance.

It is easy to see that
\begin{equation}
    {\displaystyle JSD(p||q)=H(m)-{\frac {1}{2}}{\bigg (}H(p)+H(q){\bigg )}}.
\end{equation}


Another quantity related to the complexity of the system is ''disequilibrium,'' denoted by $D$, which shows the deviation of a given probability distribution from an uniform one. 
The concept of statistical complexity of a system can be considered a development of the concept of entropy. In the articles \cite{lopez-ruiz_shannon_2005, catalan_features_2002, calbet_tendency_2001, chaos, LAMBERTI2004119, rosso} it is defined as
\begin{equation}
    C = H \cdot D,
\end{equation}
where $C$ is statistical complexity, $H$ is information entropy, and $D$ is a measure of the disequilibrium of the distribution relative to the uniform one. 



The measure of statistical complexity reflects the relationship between the amount of information and its disequilibrium in the system.
As a parameter $D$, according to the authors of \cite{rosso}, one can choose any metric that determines the difference between the maximum entropy and the entropy of the studied signal. The simplest example of disequilibrium is the square of the Euclidean distance in $\mathbb{R}^N$ between the original distribution and the uniform distribution, but often Jensen-Shannon divergence \cite{plane, features} is also used.

\subsection{Time entropy}
Now let us consider the information characteristics mentioned above in relation to time series. The Shannon entropy for systems with unequal probability states is defined as follows. Let $i$-th state of the system have probability $p_i = N_i / N $, where $N$ is a sample volume and $N_i$ is the amount of filling of $i$-level. Then the entropy $H(p)$ according to the formula \eqref{basic_entropy} equals
\begin{equation}\label{shannon_entropy}
   H(p) = -\sum_{i=1}^N p_i\log_2 p_i, \quad \sum_{i=1}^N p_i = 1.
\end{equation}


There are different ways to calculate probabilities $p_i$ from the time series. The simplest one is as follows. First, the maximum $x_{\max}$ and minimum $x_{\min}$ values are found for the considered time series ${x(t)}$ with $N$ data points. Then the interval ($x_{\max}-x_{\min}$) is divided into $n$ sub-intervals (levels) so that the value of the interval $\Delta x$ is not less than the confidence interval of the observations. The resulting sample is treated as a "message" and the $i$ subintervals as an "alphabet". Then we find the number $\Delta N_i$ of sample values ${x_k}$ that fall into each of the subintervals, and determine the relative population level $p_i^t$ (the probability of a value from the sample falling into a subinterval $i$, that is, the relative frequency of occurrence of the "letter" in the "message"):

\begin{equation}\label{groop}
    p_i^t = \frac{\Delta N_i}{N}, \quad \sum_{i = 1}^n \Delta N_i = N, \quad \sum_{i = 1}^n p_i^t = 1.
\end{equation}


The elementary entropy of the sampling is defined as the Shannon entropy \eqref{shannon_entropy} on a given set $p_i^t$ and normalized to the total number of states $n$ so that its values belong to the interval $[0, 1]$:
\begin{equation}\label{entrTNorm}
   \displaystyle H(p^t) = \frac{\displaystyle -\sum_{i=1}^n p_i^t \log_2 p_i^t}{\log_2 n}.
\end{equation}


This approach is known as the first sampling entropy \cite{SampleEntropy} and is used, for example, in \cite{US_patent} to detect the hydroacoustic signal emitted by an underwater source. 


On the other hand, the second sampling entropy can be defined as 
\begin{equation}\label{p0}
  \displaystyle  H(p^0) = \frac{\displaystyle -\sum_{i=1}^N p_i^0 \log_2 p_i^0}{\log_2 N}, \quad p_i^0=\frac{x(t_i)}{\displaystyle\sum_{k=1}^N x(t_k)}.
\end{equation}

In this case the signal samples themselves are considered as "letters", which are distributed across the time axis in contrast to amplitude axis from \eqref{entrTNorm} and the "alphabet" is the whole set of the amplitudes.

\subsection{Spectral entropy}

In addition to the time domain, the entropy can also be calculated based on the representation of the signal in the frequency domain, i.e. $p_i$ can be calculated using the spectrum of the signal. Spectral entropy is a quantitative assessment of the spectral complexity of the signal in the frequency domain from the energy point of view.

Consider time series $x(t)$ and its spectral decomposition in the frequency domain $X(f_i)$ with $N_{fft}$ frequency components, obtained using Fast Fourier Transform (FFT). The spectral power density is estimated as follows:
\begin{equation}\label{spectral_s}
  s(f_i) = \frac{1}{N_{fft}}|X(f_i)|^2.  
\end{equation}
Then the probability distribution of the spectral power density $p^s = \{p_1, p_2, . . . . , p_{N_{fft}}\}$ can be written in the form
\begin{equation}\label{spectral_prob}
 \displaystyle   p_i^s = \frac{s(f_i)}{\displaystyle\sum_k^{N_{fft}} s(f_k)}, \quad i = 1, \dots,~N_{fft}, 
\end{equation}
where $s(f_i)$ is the spectral energy for the spectral component with frequency $f_i$, $p^s_i$ is the corresponding probability density, $N_{fft}$ is the number of spectral components in the FFT, and the upper index $s$ shows that the distribution refers to the signal spectrum. The resulting function is a spectrum distribution density function.

Finally, the spectral entropy can be determined
by the equation \eqref{shannon_entropy} and normalized by the size of the spectrum:

\begin{equation}\label{EntNormf}
 \displaystyle   H(p^s) = \frac{ \displaystyle -\sum_k^{N_{fft}} p_k^s\log_2 p_k^s}{\log_2 N_{fft}}.
\end{equation}

Moreover, the Matlab mathematical package has recently implemented the Spectral Information Divergence ($SID$) method \cite{SID}, which calculates according to formula \eqref{symmer_Kullback} the similarity of two signals based on the divergence between the probability distributions of their spectra:
\begin{equation}\label{SID}
    SID(r, t) = \sum_{i} p_i \log\frac{p_i}{q_i} + \sum_{i} q_i \log\frac{q_i}{p_i},
\end{equation}
where $r$ and $t$ are the reference and test spectra, respectively, and the values of the probability distribution $p$ and $q$ for these spectra are determined according to \eqref{spectral_prob}.

\section{Entropy variation and related information criteria}\label{EntVariation}

\subsection{One-Dimensional Distributions}
The purpose of this section is to obtain the most appropriate formulas for calculating the information criteria that are responsible for the differences between distributions. Let us consider an entropy variation with respect to variation of the probability distribution. The following lemma is valid. 
\begin{Lemma}\label{lemma1}
For small variations $\delta q_i=p_i-q_i$ of the discrete distribution $q_i$ at $i=1,...,N$, so that $p_i$ is also some discrete distribution, the decomposition of entropy variation $\delta H$ in case of series convergence by powers $\displaystyle{\delta q_i}$ has a form
\begin{equation}\label{var_entrop}
    \displaystyle \delta H=H(q+\delta q)-H(q)=LH(p||q)-\frac{D(p,q) N}{2\ln 2}+o\left(\displaystyle\left(\delta q\right)^2\right),
\end{equation}
The first summand of the entropy variation decomposition $\delta H$ is the difference of cross-entropy and entropy, and the second one depends on the weighted squares of the variation of the distribution:
\begin{equation}\label{LH}
LH(p||q)=H(p,q)-H(q),
\end{equation}
\begin{equation}\label{disequ}
    \displaystyle D(p,q) = \frac{1}{N} \sum_{i=1}^N \frac{ \left(\delta q_i \right)^2}{q_i}=\frac{1}{N} \sum_{i=1}^N\frac{ \left(p_i - q_i \right)^2}{q_i}=\frac{1}{N} \left(\sum_{i=1}^N q_i\left(\frac{p_i}{q_i}-1\right)^2\right).
\end{equation}
\end{Lemma}
The proof of the Lemma \ref{lemma1} is given in Appendix A.
\begin{Remark}\label{Disequilibrium}
If $q$ is the uniform distribution, i.e. $q_i=1/N$ for $i=1,...,N$, when
$$LH(p||q)=H(p,q)-H(q)=0,$$
and the disequilibrium $D=D(p,q)$ is proportional to variance of the distribution $p$ relative to the uniform one and is equal to
\begin{equation}\label{disequ_1}
    \displaystyle D_{SQ}(p,q)= \sum_{i=1}^N\left(p_i - \frac{1}{N} \right)^2=\sum_{i=1}^N p_i^2 - \frac{1}{N} .
\end{equation}
\end{Remark}

Equation \eqref{disequ_1} coincides with disequilibrium definition from \cite{lopez-ruiz_shannon_2005}. 

According to Lemma \ref{lemma1} and Remark \ref{Disequilibrium} we can introduce a new definition.

\begin{Definition}
In the case when $q$ is the uniform distribution statistical complexity is proportional to the first non-zero member of the raw of square entropy variation, namely 
\begin{equation}\label{comp}
   C_{SQ}= H(p)\cdot D_{SQ}(p,q)=\left(-\sum_i^N p_i\log_2 p_i \right)\cdot \left(\sum_{i=1}^N \left(p_i - \frac{1}{N} \right)^2\right).
\end{equation}
\end{Definition}

\begin{Remark}
In general case statistical complexity is defined as
\begin{equation}\label{compG}
   C\sim \delta (H)^2\sim H\delta H=H(H_{max}-H),
\end{equation}
where $H_{max}$ is an entropy maximum.
\end{Remark}

It follows from Remark \ref{Disequilibrium} that the disequilibrium (\ref{disequ_1}) and the complexity (\ref{comp}) must be applied when evaluating and comparing signals with a background noise having a spectral distribution close to a uniform one. 

The formula for disequilibrium \eqref{disequ_1} proposed in \cite{rosso} has been derived from entropy variation, but most papers use Jensen-Shannon divergence \cite{features} as disequilibrium:
\begin{equation}\label{DJSD}
D_{JSD}=JSD(p||q),
\end{equation}
where $q_i=1/N$. Statistical complexity, correspondingly, is expressed as
\begin{equation}\label{compJSD}
    C_{JSD}(p) = H(p)\cdot JSD(p||q).
\end{equation}
Further in the article the comparison of the complexity graphs calculated for these two values $D_{SQ}$ and $D_{JSD}$ will be presented.

Considering the signal distribution in the frequency and time domains, we can notice that the spectral distribution does not require any additional estimation of the signal variance. Whereas when calculating entropy in the time domain, variance estimation is required, since for white noise (Gauss distribution) the following formula is valid
\begin{equation}\label{gauss_ent}
H_w=-\int_{-\infty}^{+\infty} \rho^t(x)\log_2(\rho^t(x)) dx=\log_2\left(\sqrt{2\pi e}\sigma \right),
\end{equation}
where $\displaystyle \rho^t(x)=\frac{1}{\sqrt{2\pi}\sigma}\exp(-(x-\mu)^2/(2\sigma^2))$ is the Gaussian distribution.
\begin{Remark}\label{rem2}
For the case of two continuous distributions $(p,q)$ disequilibrium $D(p,q)$ is equivalent to $f$-divergence \cite{fdivergence} with quadratic function $f$:
\begin{equation}\label{cont}
\displaystyle D(p,q)=\int \rho_q(x) \left(\frac{\rho_p(x)}{\rho_q(x)}\right)^2 dx,  
\end{equation}
if an integral exists.
\end{Remark}
\begin{Remark}\label{rem3}
In the case of two Gaussian distributions with parameters $(\mu_p,\sigma_p)$ and $(\mu_q,\sigma_q)$ formulas \eqref{LH} and \eqref{disequ} from Lemma \ref{lemma1} take the form
\begin{equation}\label{DGauss}
\displaystyle D(p,q)=\frac{\sigma_q^2}{\sigma_p\sqrt{2\sigma_q^2-\sigma_p^2}}\exp\left(\frac{(\mu_p-\mu_q)^2}{2\sigma_q^2-\sigma_p^2}\right)-1,  
\end{equation}
\begin{equation}\label{HGauss}
\displaystyle LH(p||q)=H(p,q)-H(q)=\frac{(\mu_p-\mu_q)^2}{2\sigma_q^2}+\frac{1}{2}\left(\frac{\sigma_p^2}{\sigma_q^2}-1\right).
\end{equation}
\end{Remark}

The formulas \eqref{DGauss} and \eqref{HGauss} are obtained by calculating the integrals \eqref{cont} and \eqref{gauss_ent} for continuous distributions, but are still applicable to discrete ones.


The problem in question is determining the most informative methods for calculating entropy and other information criteria. In our opinion, the answer can be obtained only in the presence of additional knowledge about the phenomenon under study. Indeed, measuring the amplitude of the signal $x(t)$ initially there is only knowledge of the time series samples $x(t_i)$, namely, we know the values of amplitudes in the increasing sequence of time samples $t_i$, $i=1,...,N$. Setting the distribution density using the formula (\ref{groop}) itself defines some random variable.

Let us calculate the entropy $H(p^0)$, applying grouping (\ref{groop}) and considering the fact that entropy does not change with changing of the summation order. Then the chain of equations is valid:
\begin{equation}\label{connection}
\begin{array}{cc}
\displaystyle H(p^0)=-\sum_{i=1}^N p_i^0\log_2 p_i^0\approx-\sum_{j=1}^n p_j^1\log_2 \frac{p_j^1}{\Delta N_j}=\log_2 N+H(p^1)-H(p^1,p^t)=\\
\displaystyle=\log_2 N-D_{KL}(p^1||p^t),
\end{array}
\end{equation}
where 
\begin{equation}\label{p1}
\displaystyle p_j^1=\sum_{i\in I_j} p_i^0,~~\displaystyle I_j=\left\{i\in(1,...,N): p_i^0\in\left[\frac{j-1}{n}\max_k p_k^0,\frac{j}{n}\max_k p_k^0\right]\right\},
\end{equation} so that
\begin{equation}\label{pt}
\displaystyle p_j^t=\frac{\Delta N_j}{N},~~\displaystyle \Delta N_j = \sum_{i\in I_j} 1.
\end{equation}

Thus, we have obtained that
\begin{equation}\label{connection1}
\displaystyle H(p^0)\approx\log_2 N-D_{KL}(p^1||p^t).
\end{equation}

If now $\displaystyle p_j^t=\frac{\Delta N_j}{N}=\frac{1}{n}$ is a uniform distribution, then (\ref{connection1}) takes the form

\begin{equation}\label{connection2}
\displaystyle H(p^0)=\log_2 N-\log_2 n+H(p^1).
\end{equation}

The formula (\ref{connection1}) shows the relationship between the entropy calculated from time samples and the Kullback-Leibler distance between the distributions obtained by alphabetical grouping along the amplitude and time coordinate axes. Therefore the next statement is valid.
\begin{Corollary}
The value of $D_{KL}(p^1||p^t)$ is approximately a constant value, independent of the method of grouping when the number of letters of the alphabet is large enough when their number is in some interval of values.
\end{Corollary}


The following observation considering distribution $p^1$ is true.
\begin{Remark}\label{Ergod}
If the sequence of samples $x(t_i)$ has the property of ergodicity, and signal is represented with white noise, then if the number of letters of the alphabet from (\ref{p1}) is large enough, the density $p^1$ is close to Gaussian.
\end{Remark}

Remark \ref{Ergod} allows to estimate $H(p^0)$ using equation \eqref{gauss_ent} for Gaussian $p^1$.

\subsection{Two-Dimensional Distributions}

The next step in constructing information criteria is to study a two-dimensional discrete random variable $(X,Y)$ with a distribution law $p_{ij}=p(X=x_i,Y=y_j)=p_1(X=x_i)p_2(Y=y_j)=p_{1,i}p_{2,j}$ for $~i=1,2...,N,~j=1,2...,K$. 
The entropy of such a distribution is given as follows
\begin{equation}\label{two_dim_distribution}
    \displaystyle \mathrm {H} (p)=-\sum_{i=1}^{N}\sum_{j=1}^{K}p_{ij}\log_2 p_{ij}=-\sum_{i=1}^{N}\sum_{j=1}^{K}p_{1,i}p_{2,j}\log_2 p_{1,i}p_{2,j}=\mathrm {H} (p_1)+\mathrm {H} (p_2).
\end{equation}
The disequilibrium $D(p,q)$ of the distribution $p$ with respect to the two-dimensional distribution $\displaystyle q_{kl}=q_{1,l}q_{2,k}=\frac{1}{NK}$ for $l=1,2...,N,~k=1,2...,K$,where $q_1,q_2$ are uniform distributions, is found based on the multivariate analog of Remark \ref{Disequilibrium}
\begin{equation}\label{2DD1}
\begin{array}{c}
    \displaystyle D(p,q) = \frac{1}{NK} \sum_{i=1}^N \sum_{j=1}^K\frac{ \left(p_{ij} - q_{ij} \right)^2}{q_{ij}}= \\
    \displaystyle \frac{1}{NK} \sum_{i=1}^N \sum_{j=1}^K\frac{p_{ij}^2}{q_{ij}}-\frac{1}{NK}= \frac{1}{NK} \sum_{i=1}^N \sum_{j=1}^KNKp_{ij}^2-\frac{1}{NK}
    \displaystyle= \sum_{i=1}^N p_{1,i}^2 \sum_{j=1}^K p_{2,j}^2-\frac{1}{NK}.
\end{array}
\end{equation}
In the resulting expression, we can replace each of the sums of the first summand with the corresponding disequilibrium from Remark \ref{Disequilibrium}
\begin{equation}\label{2DD2}
\begin{array}{c}
    \displaystyle D(p,q)=\left(D(p_1,q_1)+\frac{1}{N}\right)\left(D(p_2,q_2)+\frac{1}{K}\right)-\frac{1}{NK}=\\
     \displaystyle=\frac{D(p_1,q_1)}{K}+\frac{D(p_2,q_2)}{N}+D(p_1,q_1)D(p_2,q_2).
\end{array}
\end{equation}


Statistical complexity in the two-dimensional case is expressed in terms of quantities related to the one-dimensional distribution

\begin{equation}
\begin{array}{c}
    \displaystyle C=H(p)D(p,q)=D(p_1,p_1)\frac{H(p_1)}{K}+D(p_1,p_1)\frac{H(p_2)}{K}+D(p_2,p_2)\frac{H(p_1)}{N}+\\
    \displaystyle+D(p_2,p_2)\frac{H(p_2)}{N}+D(p_1,p_1)D(p_2,p_2)H(p_1)+D(p_1,p_1)D(p_2,p_2)H(p_2)=\\
    \displaystyle=\frac{C_1}{K}\left(1+\frac{H(p_2)}{H(p_1)}\right)+\frac{C_2}{N}\left(1+\frac{H(p_1)}{H(p_2)}\right)+C_1C_2\left(\frac{1}{H(p_1)}+\frac{1}{H(p_2)}\right),
    \end{array}
\end{equation}
where $C_1,~C_2$ are statistical complexities of the corresponding one-dimensional distributions. The limit transition from two-dimensional complexity $C$ to one-dimensional complexity $C_1$ is particularly interesting.

Now there are four distribution densities at our disposal, three of them $p^0$, $p^1$, $p^t$ are related to the time domain and are determined by the formulas (\ref{p0}), (\ref{p1}), (\ref{pt}), and one $p^s$, related to the signal spectrum, calculated by the formula (\ref{spectral_prob}). 


Simultaneously with the presence of four distribution densities, the following criteria are considered: normalized information entropy $H$, defined by the formulas (\ref{entrTNorm}), (\ref{EntNormf}); statistical complexity $C$, computed by (\ref{comp}) with an disequilibrium $D$ (\ref{disequ_1}); Jenson-Shannon divergence $JSD$ (\ref{J-S}); spectral information divergence $\rm{SID}$ (\ref{SID}); cross-entropy and entropy difference $LH$ (\ref{LH}).


Since the spectral density is used to compare the signal/noise mixture with white noise, i.e. with a uniform distribution, all the proposed criteria are applicable for this density. In the case of temporal distributions, the normalized information entropy $H$, which depends only on the distribution under study, and the difference of cross-entropy and entropy $LH$, calculated explicitly by the formula (\ref{HGauss}) for $\mu_p=\mu_q$, will be estimated.

On the basis of the numerical experiments performed, a conclusion will be made about the quality of the criteria used and the limits of their applicability in the presence of the noise component of the signal.

\section{Modelling and Discussion}\label{Modelling}
\subsection{About the calculation algorithm and presentation of simulation results}
In all experiments the graphs of the information characteristics are presented as a functions of time. The characteristics are calculated from the signal according to the following algorithm: 

\begin{enumerate}
    \item Digitized with sampling rate $\mathrm{F}$ audio signal is divided into short segments containing $W$ digital samples.
    \item The discrete densities $p_i$ \eqref{spectral_prob} are calculated from time or frequency domain.
    \item Information criterion is calculated using $p_i$.
    \item The sequence of values is displayed together with the signal on the time axis (each of the obtained values is extended by $W$ counts).
\end{enumerate}

Exceeding a certain threshold of information criterion indicates the appearance of a useful signal in the mixture.

The results of the signal processing according to this algorithm are presented below. For different acoustic signals a comparison of the quality of indication of the appearance of a useful signal by different information criteria at different levels of added white noise is demonstrated. In addition, the subsection \ref{C_comparison} shows a comparison of two methods for calculating statistical complexity and draws conclusion about the usefulness of both.

The first acoustic signal chosen was an audio recording of a humpback whale song recorded underwater. A large set of such recordings is available from the Watkins Marine Mammal Sound Database collected by Woods Hole Oceanographic Institution and the New Bedford Whaling Museum. The ability to separate such signals from strong sea noise may be useful for researchers-biologists for further classification and study. In addition, this signal is similar in structure to the human voice with separate words, the extraction of which can be useful, for example, in the tasks of voice activity detection and speech recognition.

In all the graphs below, the signal is marked with blue line, and the corresponding information metric -- with red line. The left vertical axis corresponds to the values of the signal amplitude, the right vertical axis corresponds to the value of the information metric. The signal is shown without added white noise for better comprehension, but the variable parameter of the standard deviation $\sigma_N$ of the noise is marked with a dashed line. All information metrics are normalized for the convenience of presentation. All calculations and visualizations were performed using Python. White noise, which was artificially added to audio recordings, was also generated numerically. 

\subsection{Time information criteria}
First, we consider the behavior of the information entropies $H(p^t)$ and $H(p^0)$, calculated from the time samples of the signal $x(t)$.

\begin{fullwidth}[leftmargin=-3cm, rightmargin= -3cm, width=\linewidth+6cm]
\begin{figure}[H]
    \centering
    \hspace{-6cm}
    \includegraphics[width = 8cm]{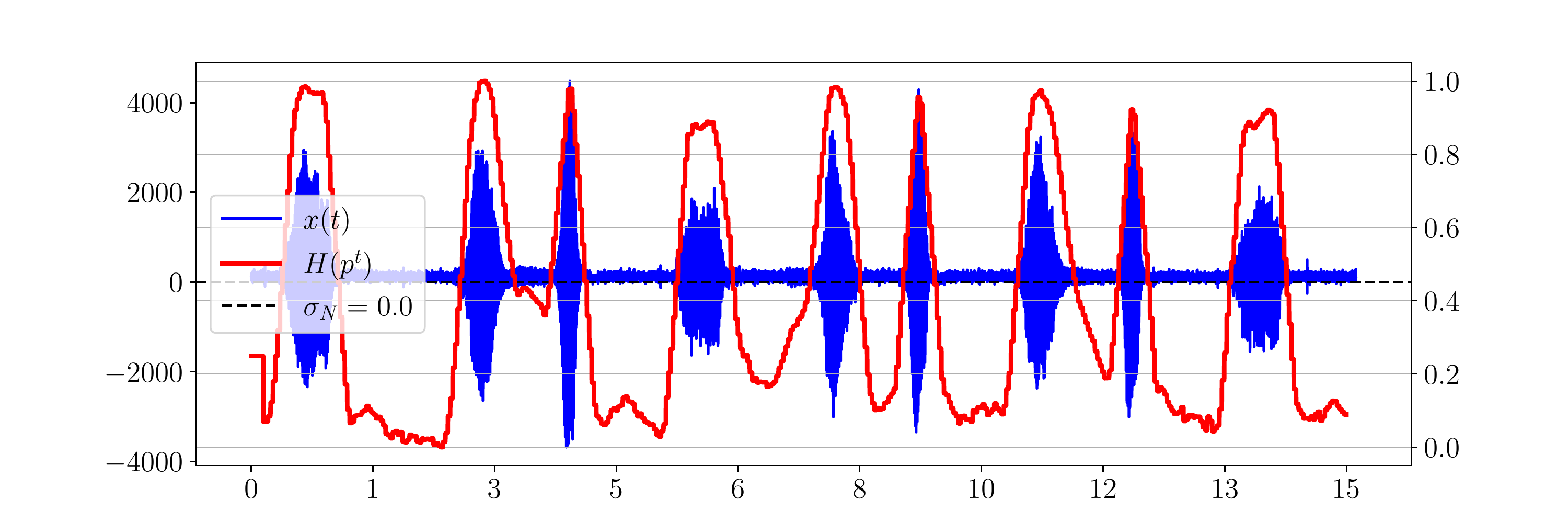}
    \hspace{-0.9cm}
    \includegraphics[width = 8cm]{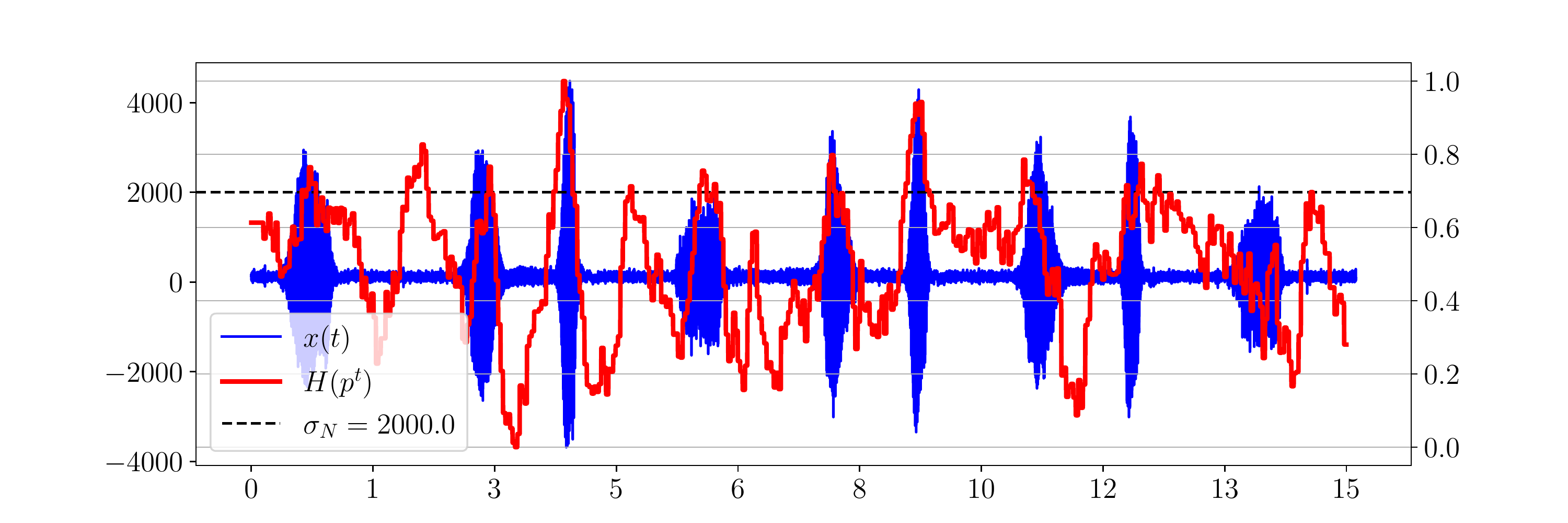}\\
    \hspace{-6cm}
    \includegraphics[width = 8cm]{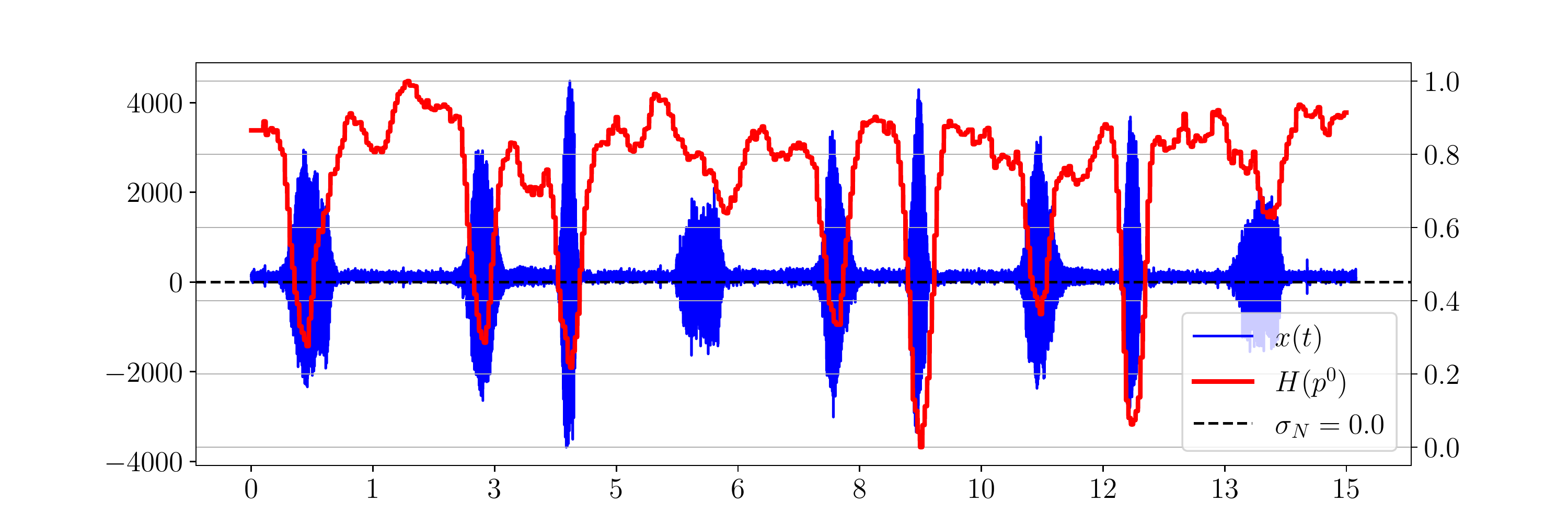}
    \hspace{-0.9cm}
    \includegraphics[width = 8cm]{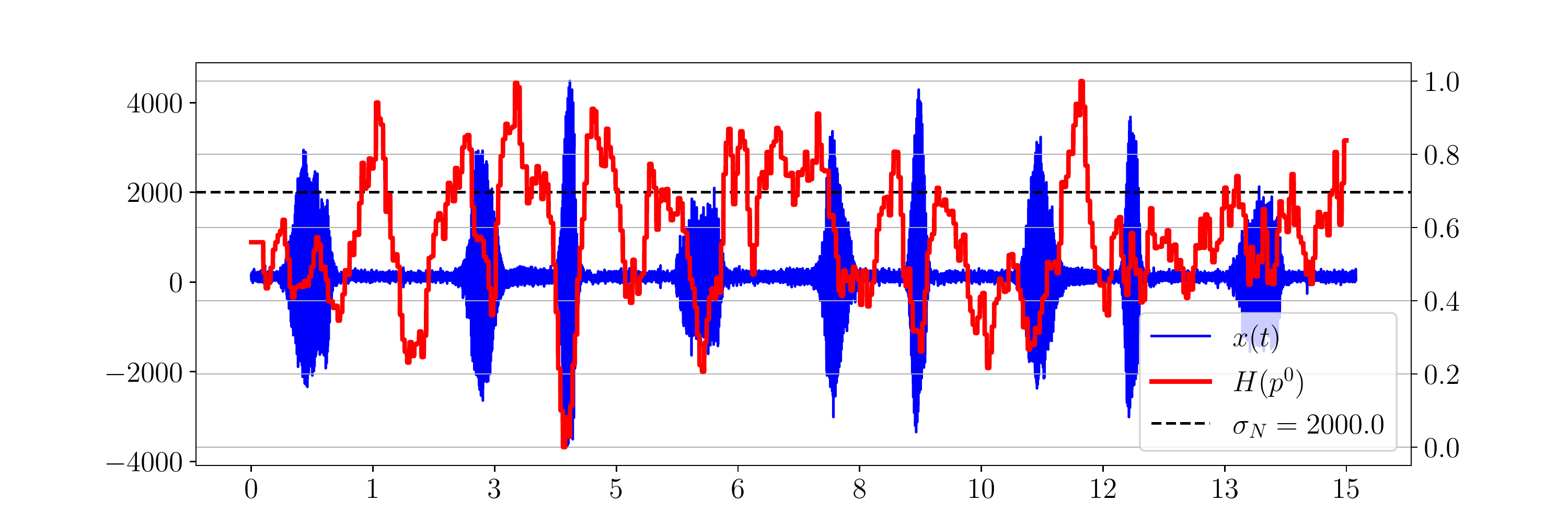}
    \caption{Graphs of $H(p^t)$ and $H(p^0)$ for different levels of added noise.}
    \label{fig:time_entropy_whale}
\end{figure}
\end{fullwidth}

Fig. \ref{fig:time_entropy_whale} shows that with noise increasing there is a serious degradation of the time entropy graph for both methods of its calculation so that for $\sigma_N = 2000$ these information criteria can no longer serve as reliable indicators of the appearance of a useful signal in the mixture. We should note an interesting feature of the behavior of $H(p^0)$ and $H(p^t)$: the first characteristic is maximum for the uniform distribution and decreases with the appearance of a useful signal in the mixture, while the second, by contrast, is minimal in the absence of signal and increases with its appearance. This obviously follows from the formulas for calculating the distributions and entropies \eqref{groop}, \eqref{p0}, \eqref{entrTNorm}.

The information characteristic $LH(p||q)$ stands out favorably from the time entropies, demonstrated above.

\begin{fullwidth}[leftmargin=-3cm, rightmargin= -3cm, width=\linewidth+6cm]
\begin{figure}[H]
    \centering
    \hspace{-6cm}
    \includegraphics[width = 8cm]{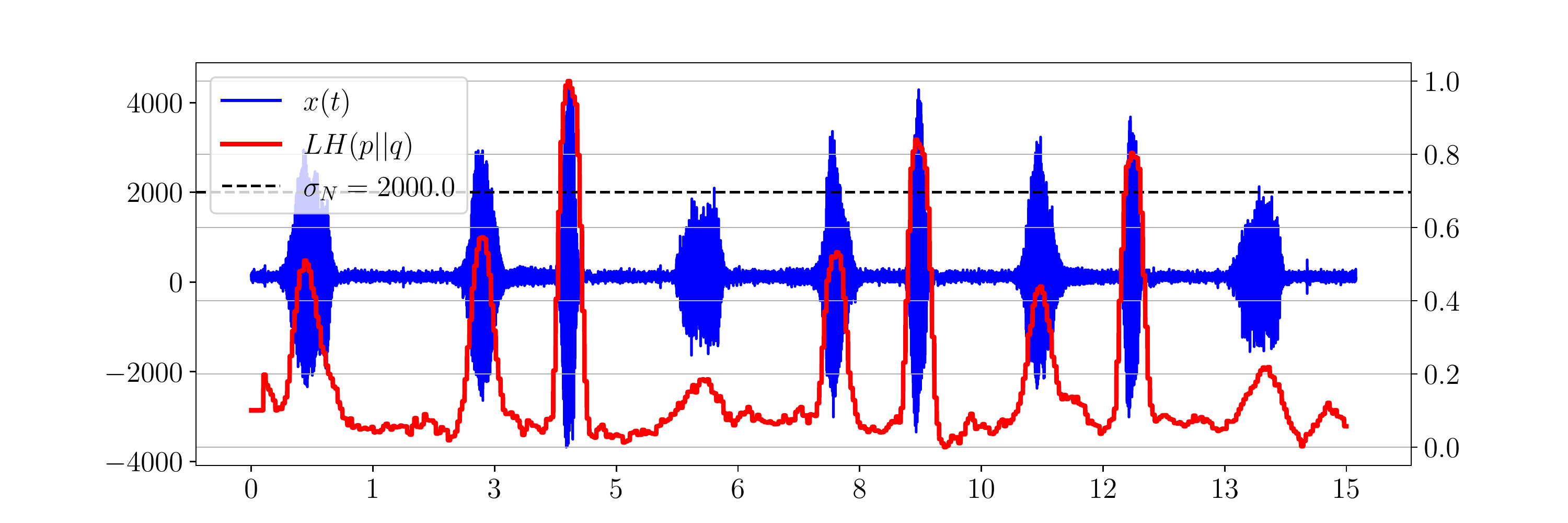}
    \hspace{-0.9cm}
    \includegraphics[width = 8cm]{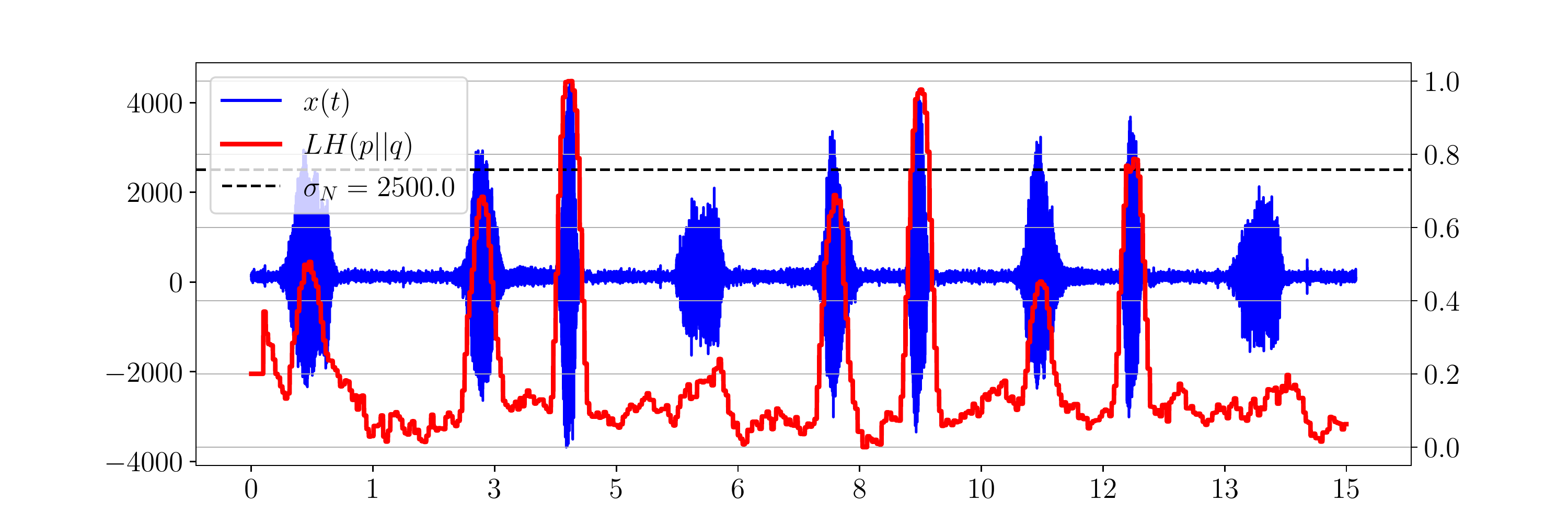}\\
    \hspace{-6cm}
    \includegraphics[width = 8cm]{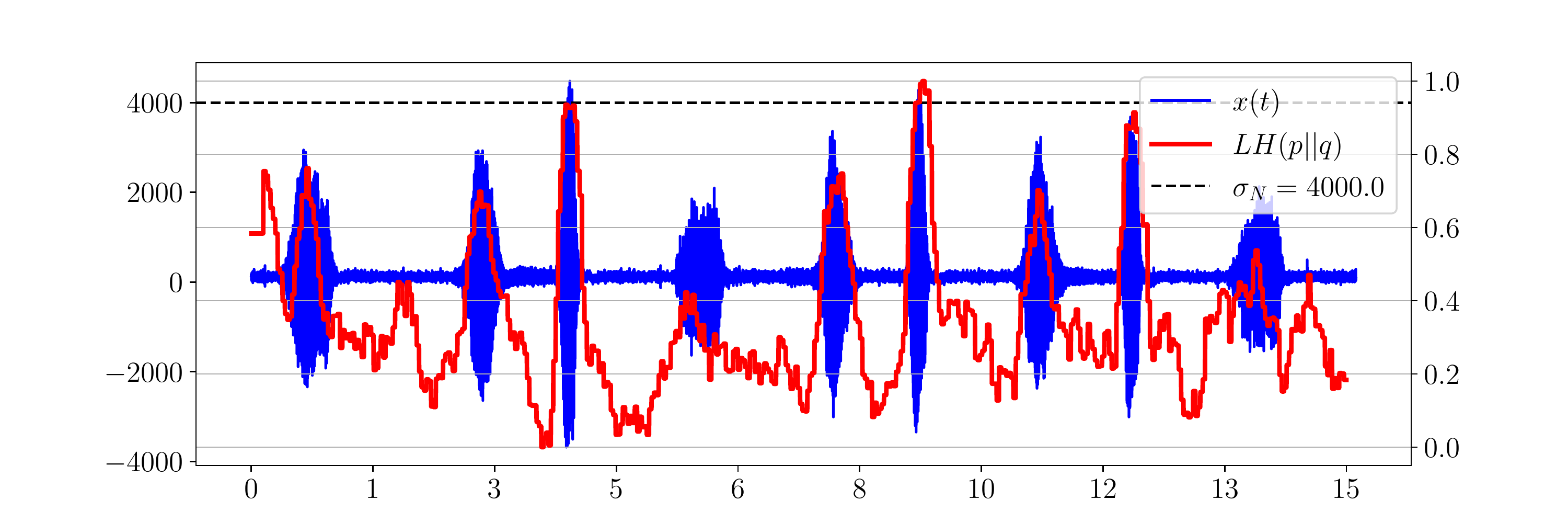}
    \hspace{-0.9cm}
    \includegraphics[width = 8cm]{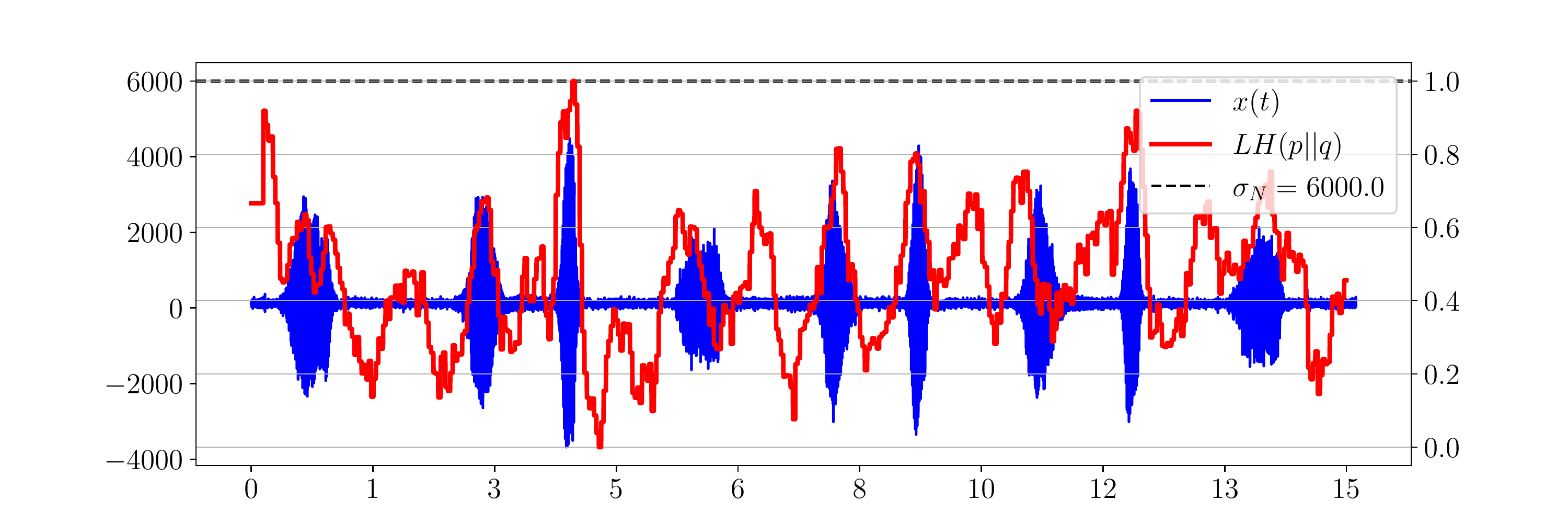}
    \caption{Graphs of $LH(p||q)$ for different levels of added noise.}
    \label{fig:LH_whale}
\end{figure}
\end{fullwidth}

In Fig. \ref{fig:LH_whale} one can see that $LH$ for the noise level $\sigma_N = 2000$ shows the appearance of a useful signal and works well enough even for the double noise value. However, it should be noted that this is true only for stationary noise, which average value does not change over time. Otherwise, this metric will react to changes in noise as well, which follows from the formula \eqref{HGauss}. Moreover, initial estimation of $\sigma_N$ is required for correct functioning of this criterion.

\subsection{Time entropy $H(p^1)$}
The time entropy $H(p^1)$ associated with another grouping of the ''alphabet'' derived from the signal samples is considered separately. 

\begin{figure}[H]
    \centering
    \includegraphics[width = 9.cm]{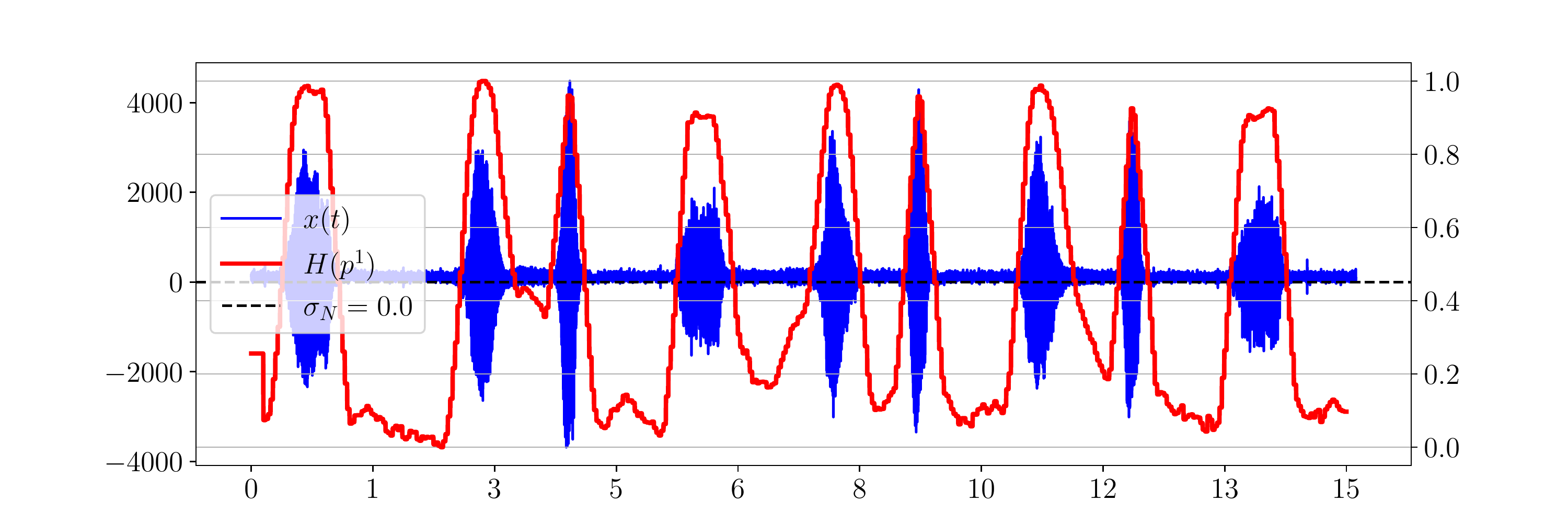}
    \hspace{-0.9cm}
    \includegraphics[width = 9.cm]{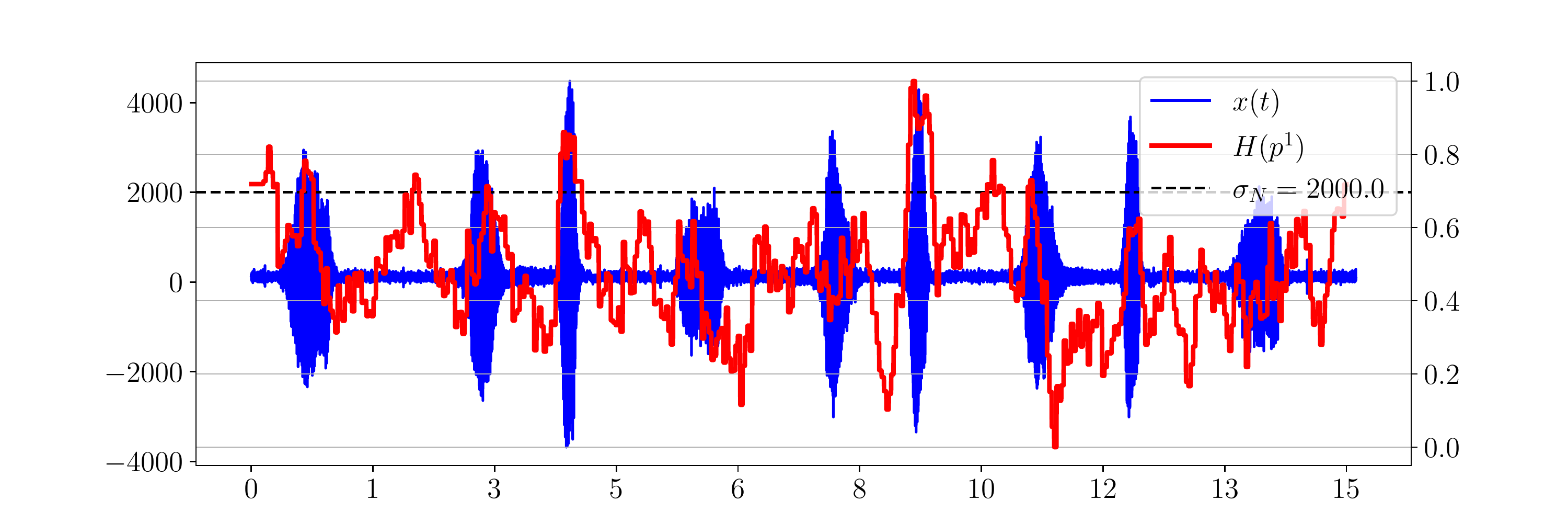}
    \caption{The entropy $H(p^1)$ for the number of letters of the alphabet equal to 64.}
    \label{fig:whale_p1_entropy}
\end{figure}
Changing the alphabet partitioning negatively affects the effectiveness of entropy in this representation:
\begin{figure}[H]
    \centering
    \includegraphics[width = 9.cm]{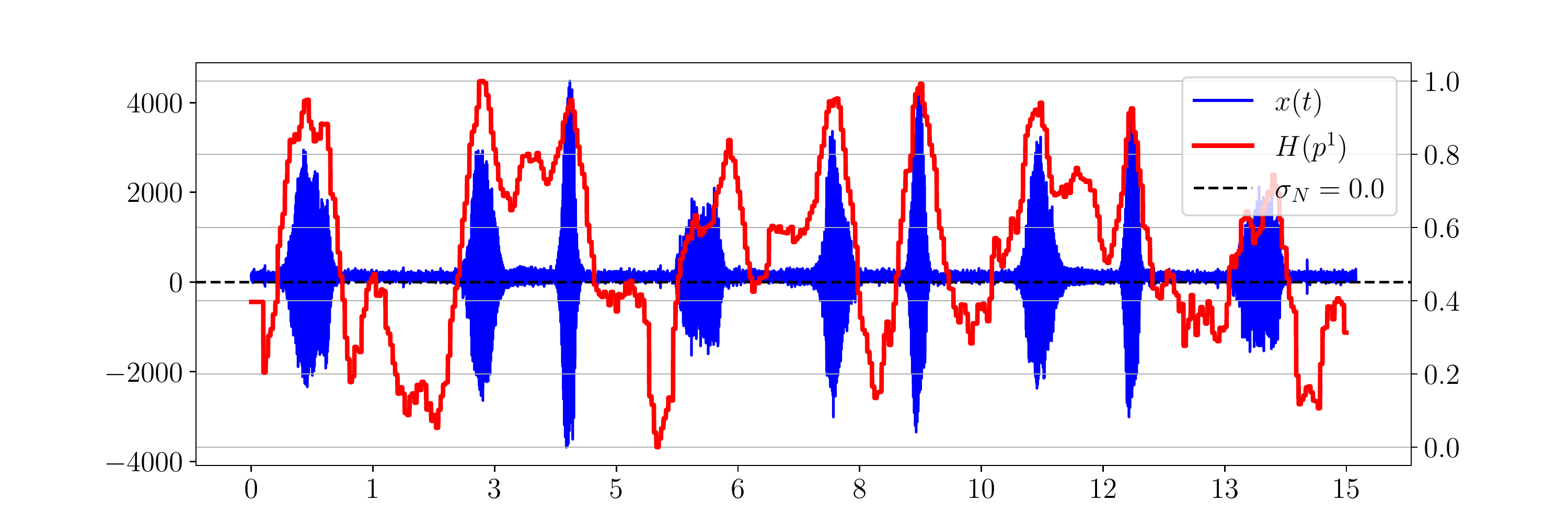}
    \hspace{-0.9cm}
    \includegraphics[width = 9.cm]{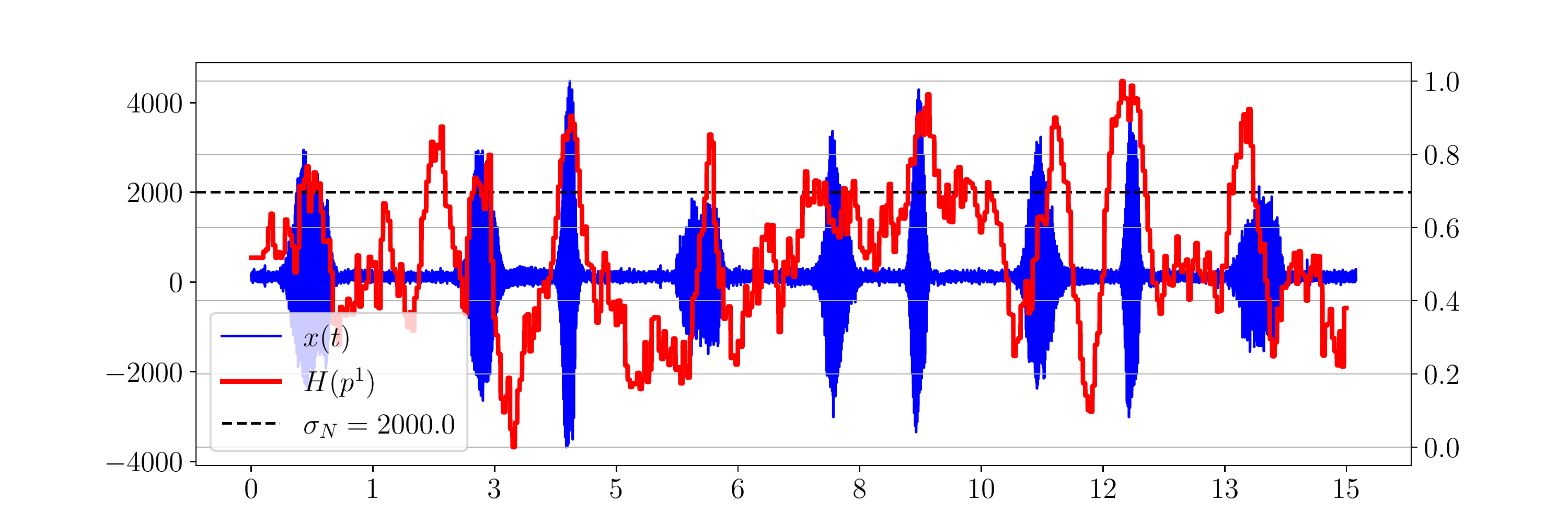}
    \caption{The entropy $H(p^1)$ for the number of letters of the alphabet equal to 8.}
    \label{fig:whale_p1new_entropy}
\end{figure}

\subsection{Spectral information criteria}
Information criteria based on the spectral distribution of $p^s$ are deprived of the disadvantages of time criteria.
\begin{fullwidth}[leftmargin=-3cm, rightmargin= -3cm, width=\linewidth+6cm]
\begin{figure}[H]
    \centering
    \hspace{-6cm}
    \includegraphics[width = 8cm]{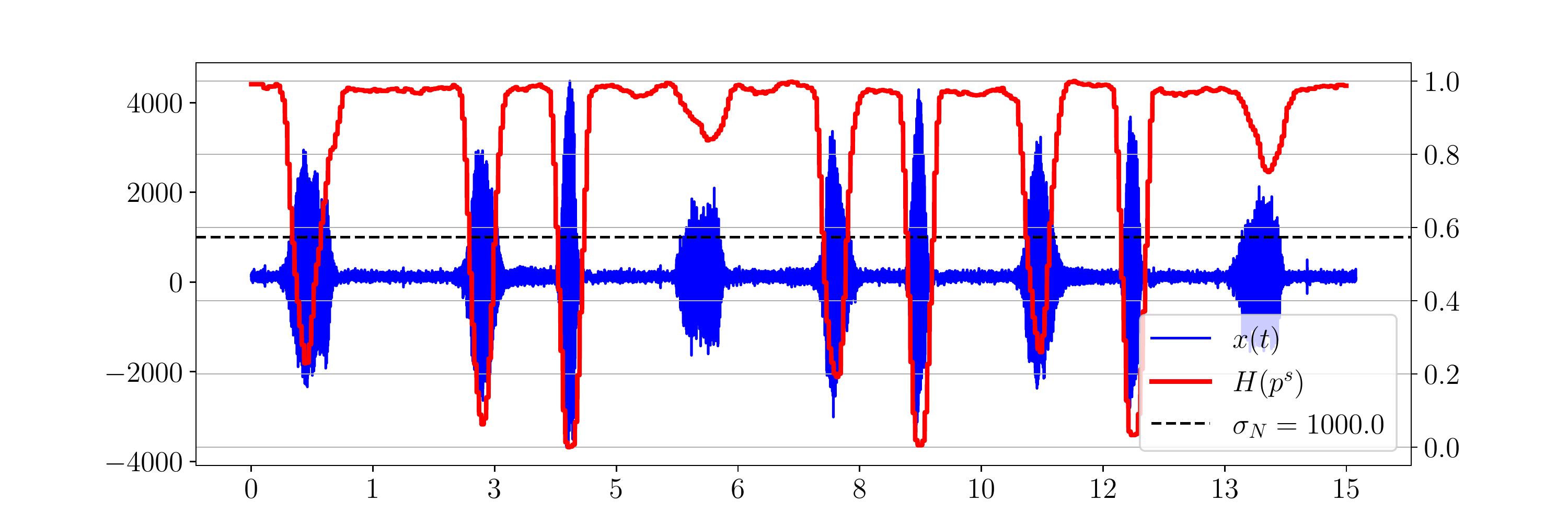}
    \hspace{-0.9cm}
    \includegraphics[width = 8cm]{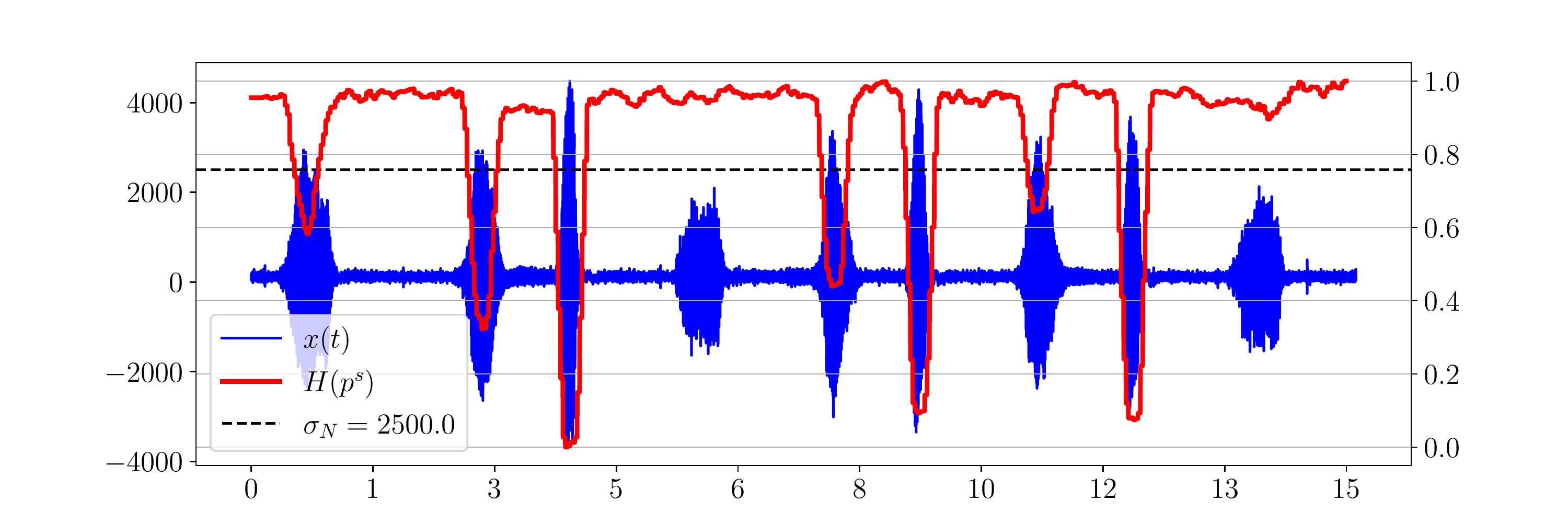}\\
    \hspace{-6cm}
    \includegraphics[width = 8cm]{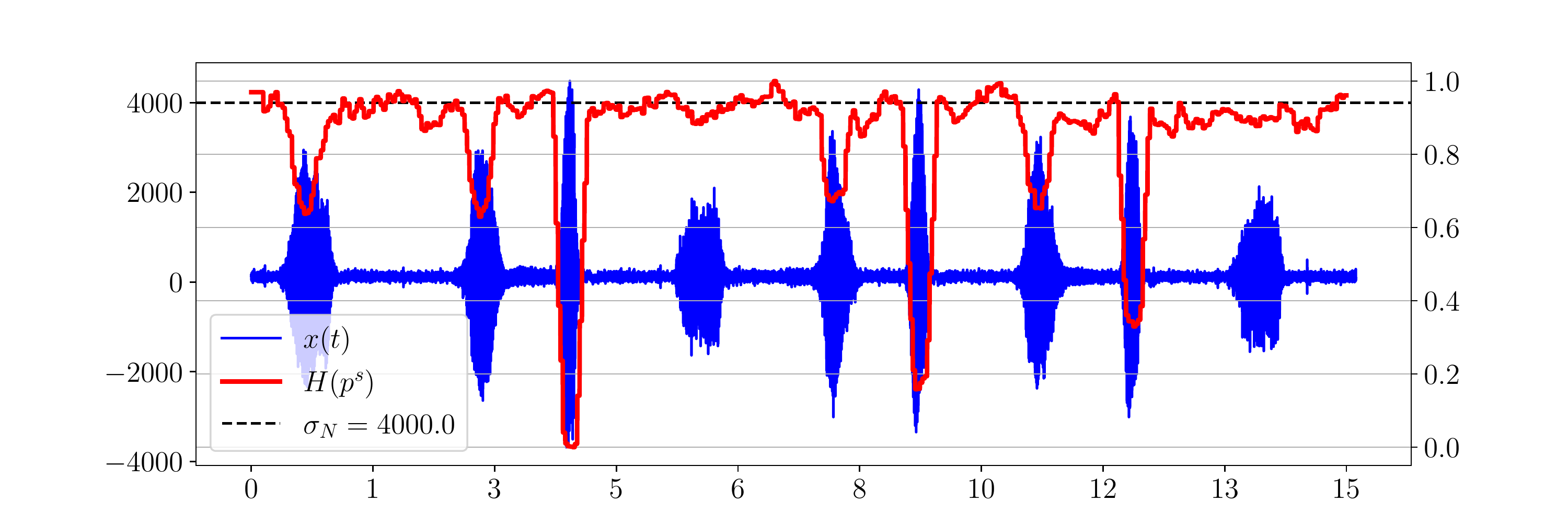}
    \hspace{-0.9cm}
    \includegraphics[width = 8cm]{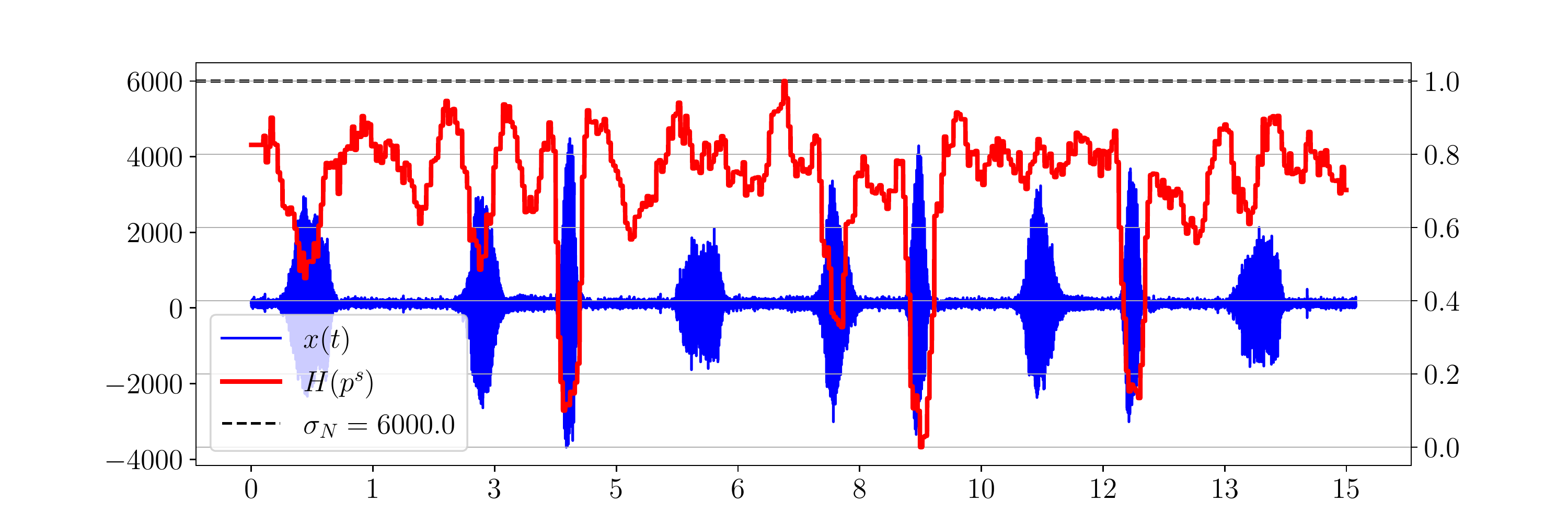}
    \caption{Spectral entropy plots for different levels of added noise.}
    \label{fig:whale_spectral_entropy}
\end{figure}
\end{fullwidth}
Fig. \ref{fig:whale_spectral_entropy} shows the dependence of the spectral entropy on time. We can see a significant improvement in the maximum allowable noise level, at which the indication of the appearance of a useful signal is still possible, with respect to the graphs in Fig. \ref{fig:LH_whale}.

The point is that the white noise in a signal in spectral representation has a quite definite uniform probability distribution, which greatly facilitates the calculation of entropy and saves us from the necessity of estimation of the variance of this noise. Moreover, even if the noise is not stationary, i.e., its parameters change over time, in a small window $W$ it can still be considered white, and the above statement is still true.

Distribution $p^s$ can be used as a basis for a number of information divergences \eqref{SID}, \eqref{DJSD}, \eqref{comp}, \eqref{compJSD}: 
\begin{fullwidth}[leftmargin=-3cm, rightmargin= -3cm, width=\linewidth+6cm]
\begin{figure}[H]
    \centering
    \hspace{-6cm}
    \includegraphics[width = 8cm]{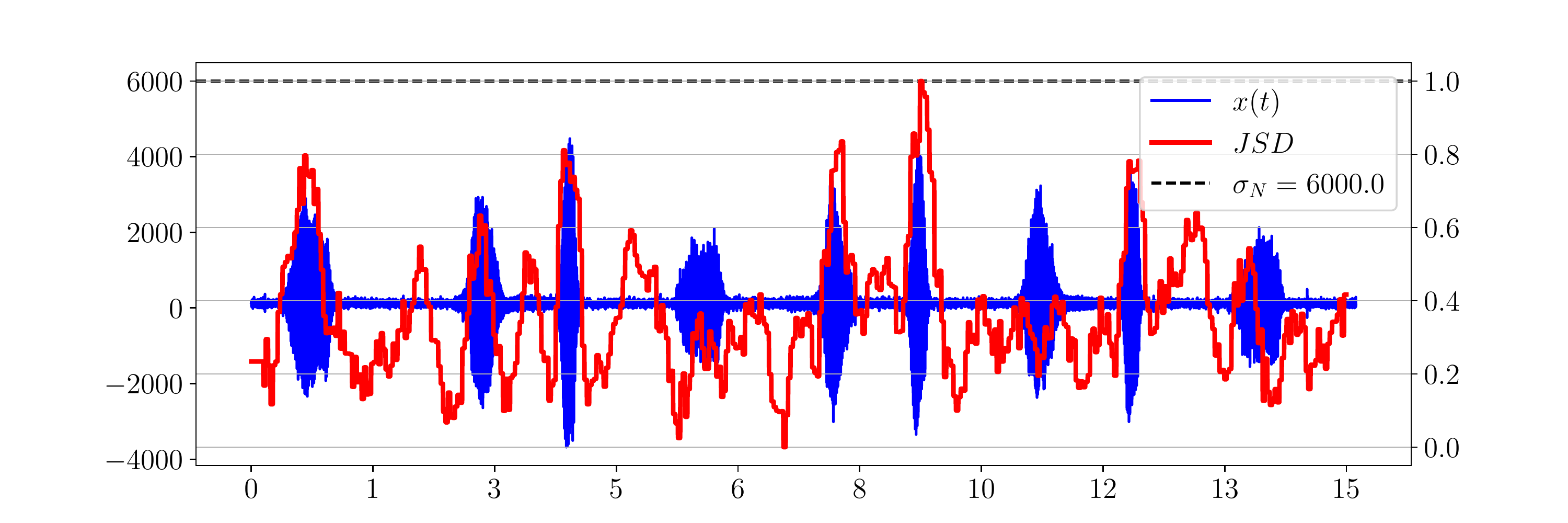}
    \hspace{-0.9cm}
    \includegraphics[width = 8cm]{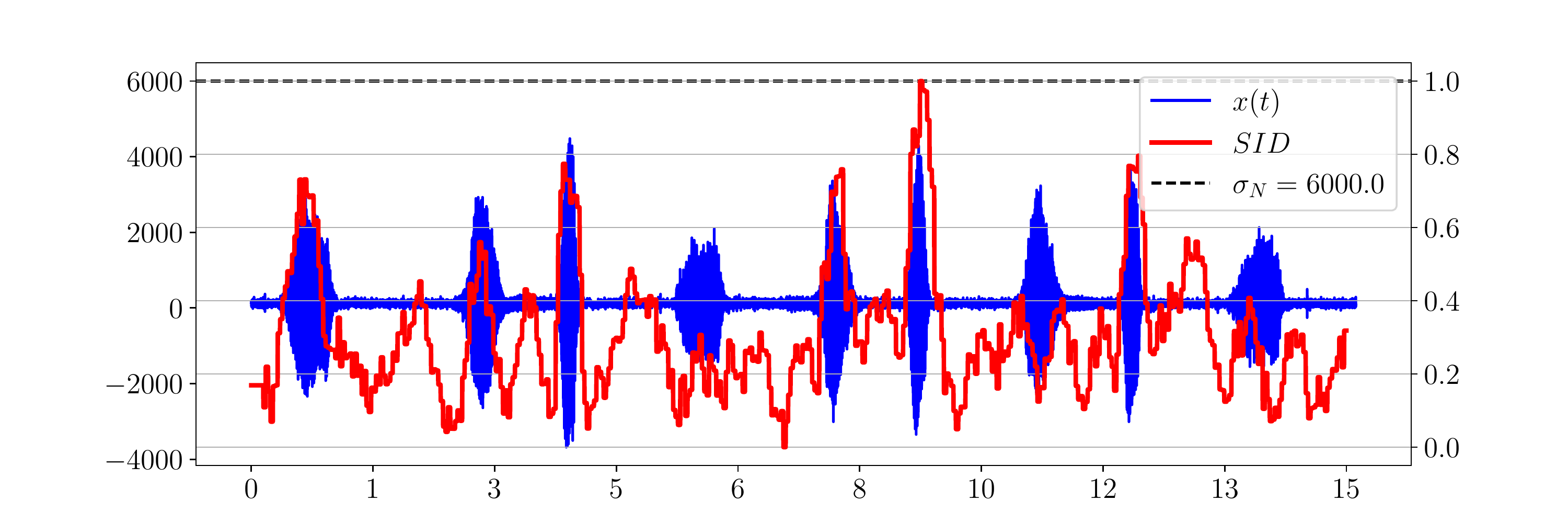}\\
    \hspace{-6cm}
    \includegraphics[width = 8cm]{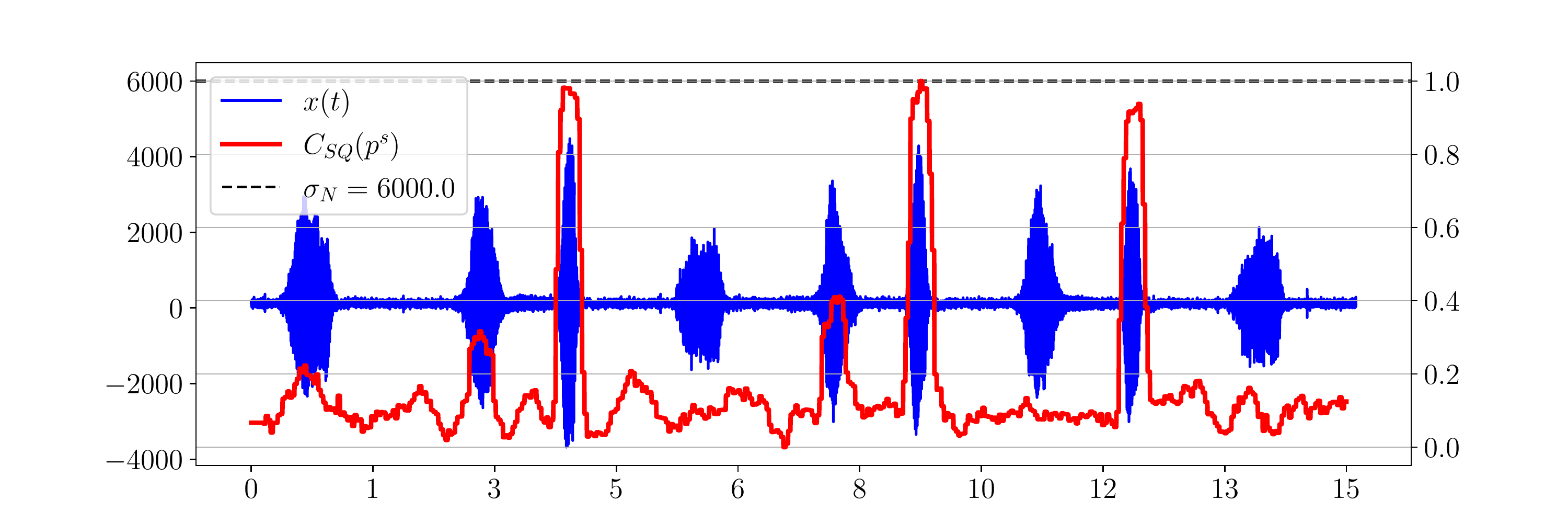}
    \hspace{-0.9cm}
    \includegraphics[width = 8cm]{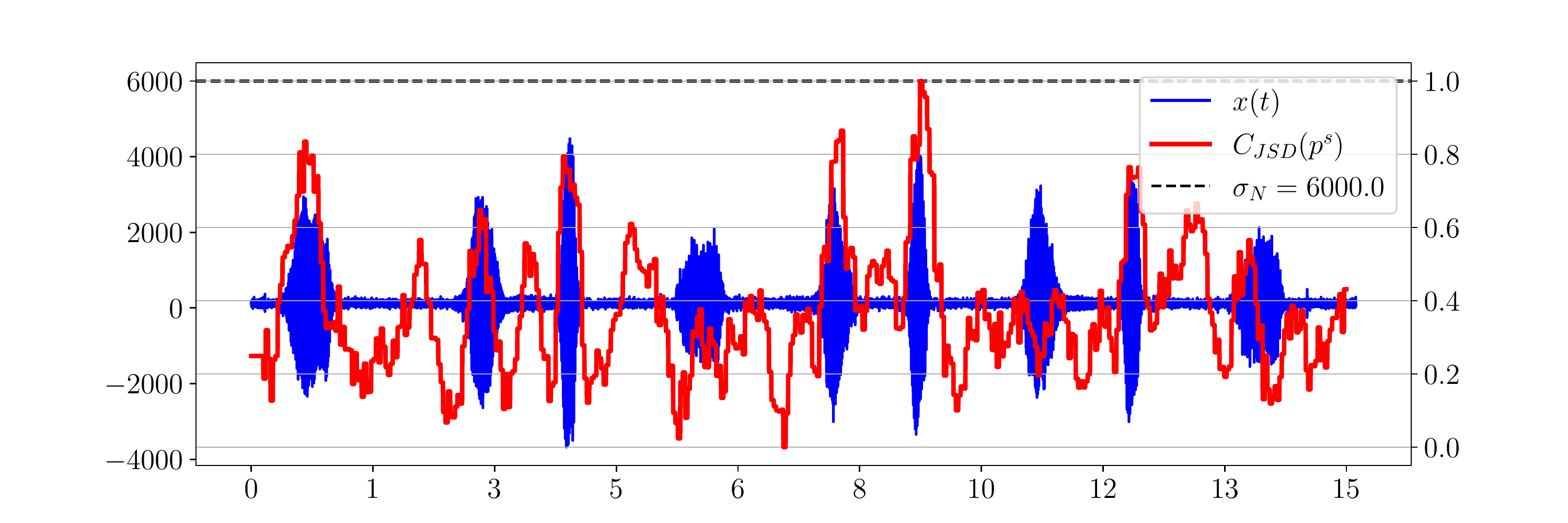}
    \caption{Information divergence plots based on spectral distribution.}
    \label{fig:whale_spectral_divs}
\end{figure}
\end{fullwidth}
Fig. \ref{fig:whale_spectral_divs} shows that the separability of information metrics decreases along with the signal-to-noise ratio (SNR). However, the statistical complexity $C_{SQ}$ performs better than all the other criteria, because it still allows to distinguish a useful signal, when other metrics behave irregularly and no longer show a significant excess of levels compared to areas without signal. Thus it is the most promising characteristic in our opinion.

\subsection{Comparison of different ways of calculating statistical complexity}\label{C_comparison}
Of separate interest is a comparison of the behavior of the statistical complexities $C_{SQ}$ and $C_{JSD}$, which essentially correspond to different ways of calculating the same value of statistical complexity.
\newpage
\begin{fullwidth}[leftmargin=-3cm, rightmargin= -3cm, width=\linewidth+6cm]
\begin{figure}[H]
    \centering
    \hspace{-6cm}
    \includegraphics[width = 8cm]{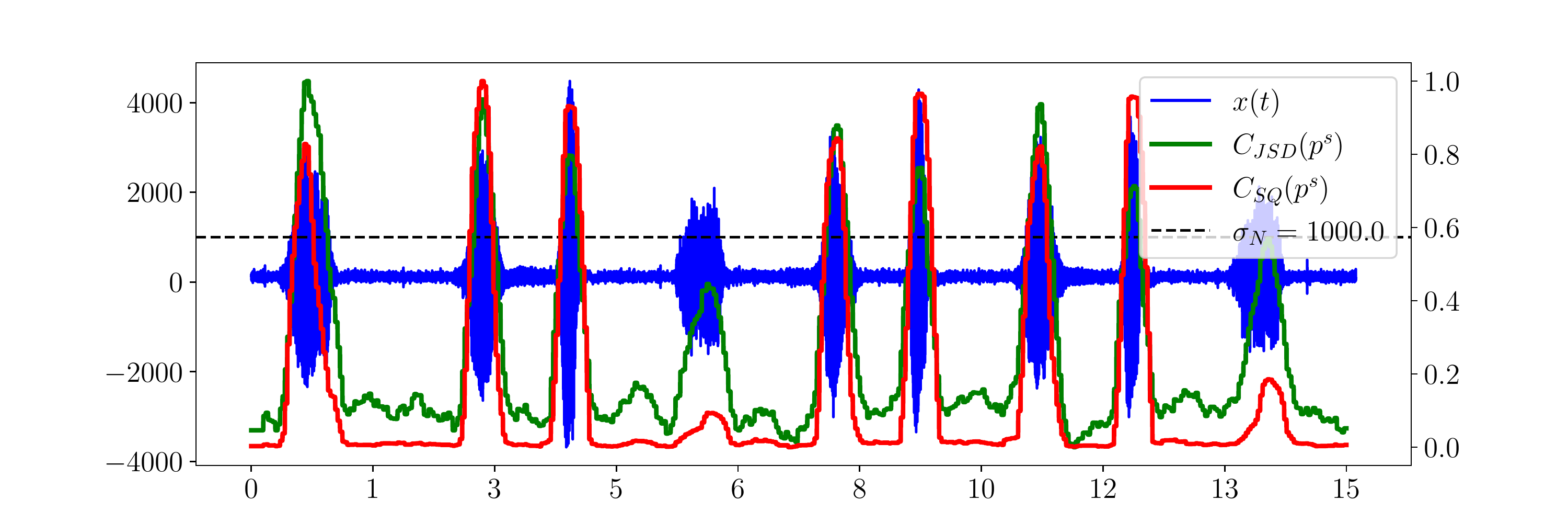}
    \hspace{-0.9cm}
    \includegraphics[width = 8cm]{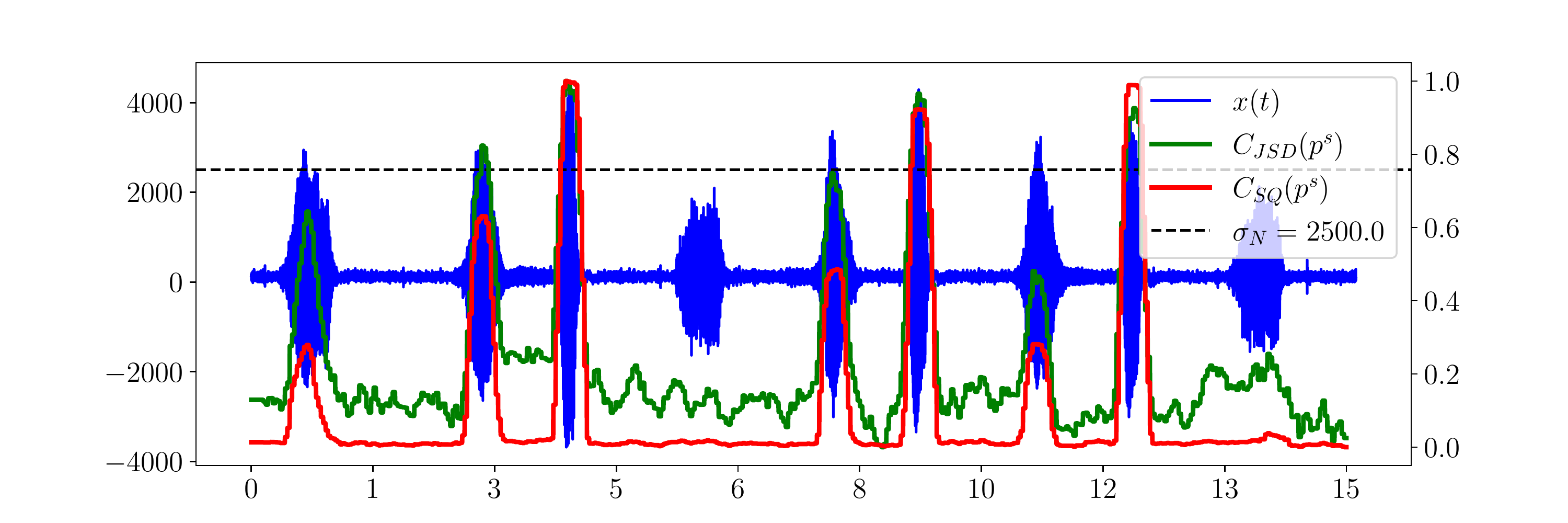}\\
    \hspace{-6cm}
    \includegraphics[width = 8cm]{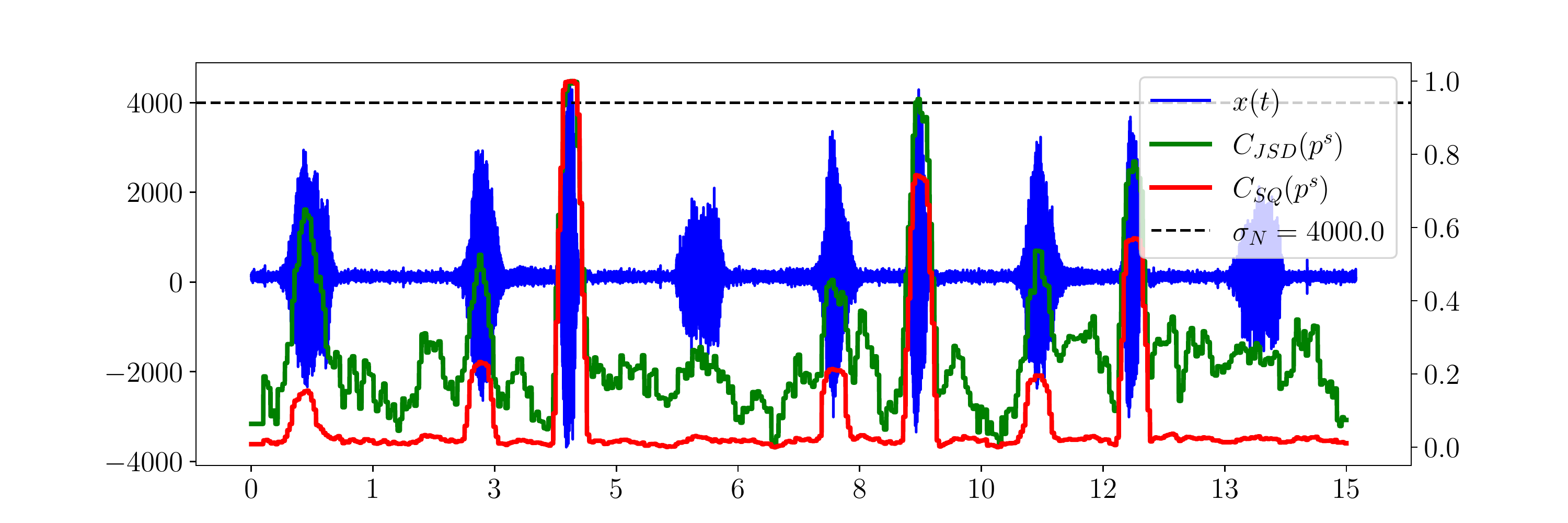}
    \hspace{-0.9cm}
    \includegraphics[width = 8cm]{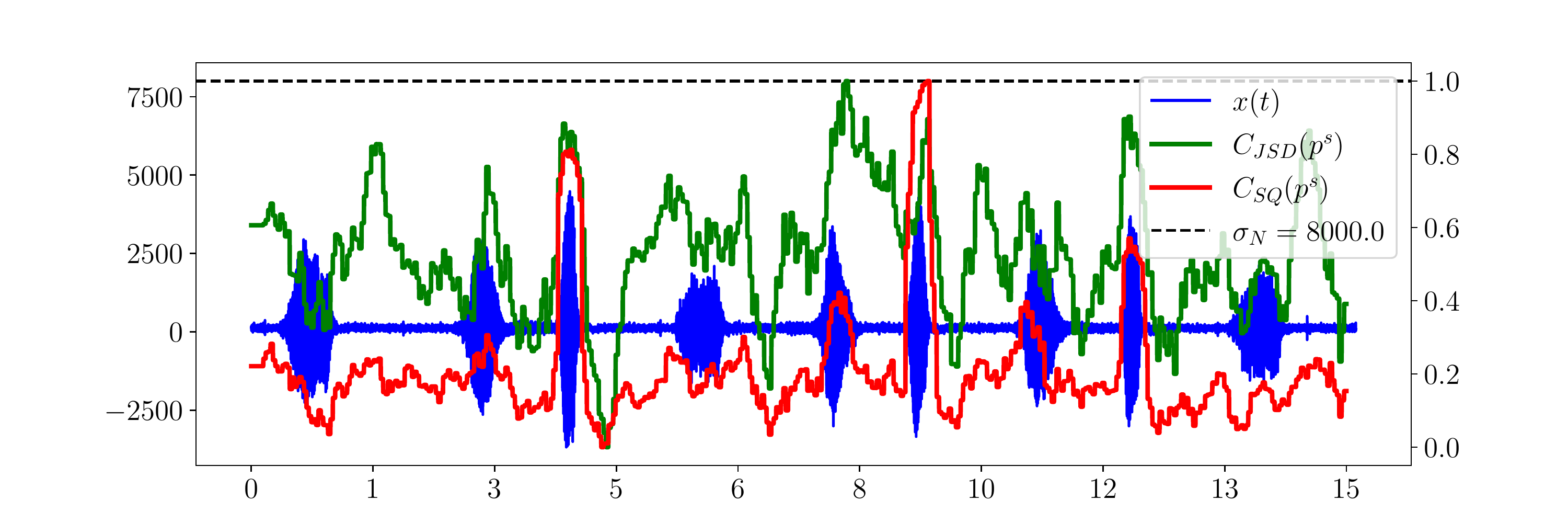}
    \caption{Comparison of statistical complexities $C_{SQ}$ and $C_{JSD}$.}
    \label{fig:whale_spectral_complexity_sq}
\end{figure}
\end{fullwidth}

We can see that $C_{SQ}$ shows a better result as an indicator of the appearance of a useful signal compared to the Jensen-Shannon divergence, which is an unexpected result, given that most articles use exactly the second method of calculation. However, it is worth noting that this metric is mostly used in classification problems utilizing the $H/C$ plane.

\subsection{Hydroacoustic signal of a marine underwater object}
The second signal is a recorded hydroacoustic signal of a marine underwater object. The study of such signals is important in military and civilian applications because it will automate the process of analyzing the hydroacoustic scene and identifying potential threats. In Fig. \ref{fig:sub_spectral_entropy} spectral entropy dependencies for different levels of added noise are shown. 
\begin{fullwidth}[leftmargin=-3cm, rightmargin= -3cm, width=\linewidth+6cm]
\begin{figure}[H]
    \centering
    \hspace{-6cm}
    \includegraphics[width = 8cm]{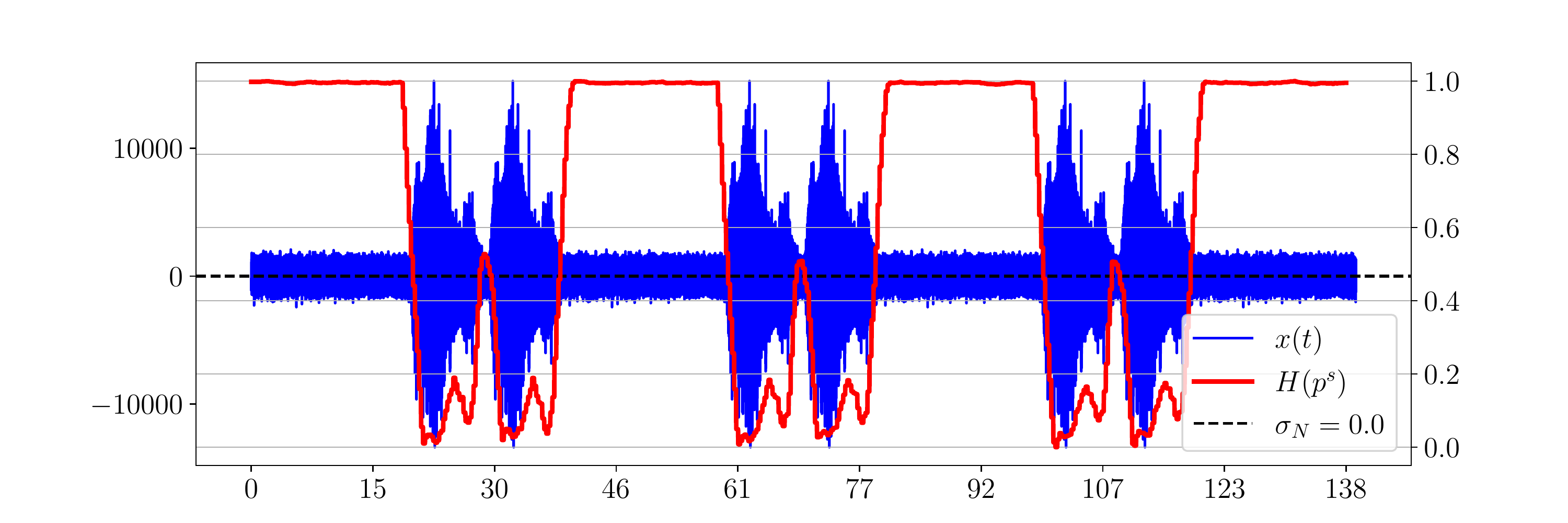}
    \hspace{-0.9cm}
    \includegraphics[width = 8cm]{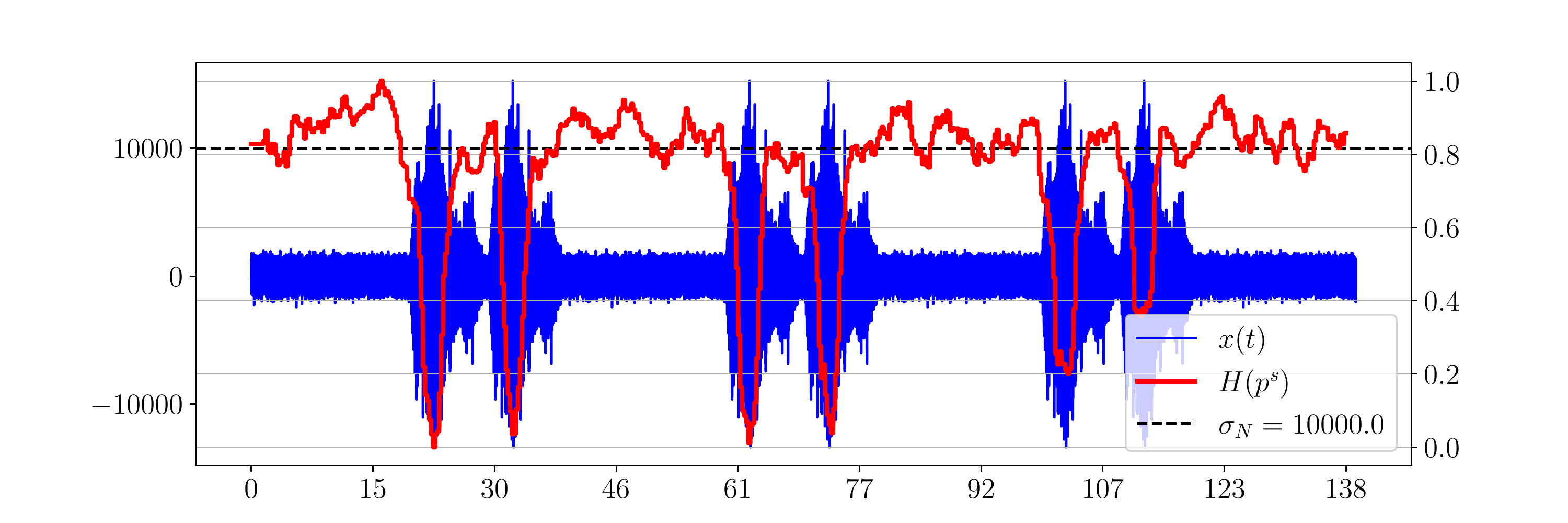}\\
    \hspace{-6cm}
    \includegraphics[width = 8cm]{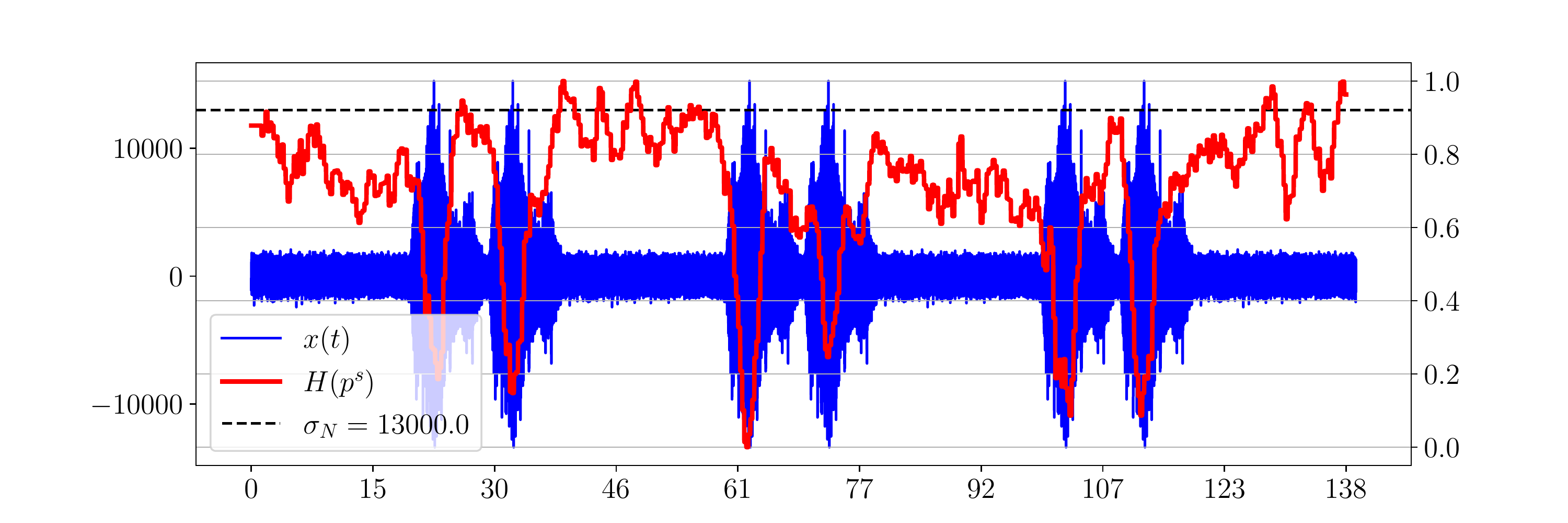}
    \hspace{-0.9cm}
     \includegraphics[width = 8cm]{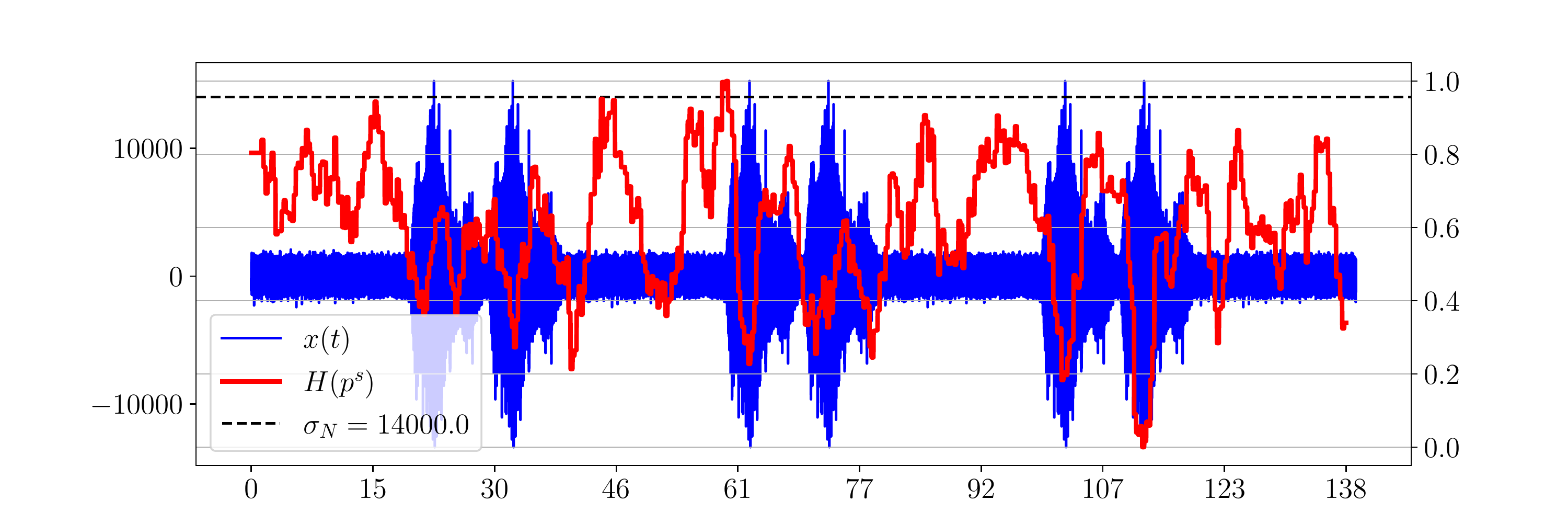}
    \caption{Spectral entropy plots for different levels of added noise.}
    \label{fig:sub_spectral_entropy}
\end{figure}
\end{fullwidth}

Fig. \ref{fig:sub_сomplexities_sq} shows the dependencies of statistical complexity for a given signal. It is worth noting that the selected information metric shows the presence of an useful signal even for a very small SNR ($\approx -15$ dB) in the last example.
\begin{fullwidth}[leftmargin=-3cm, rightmargin= -3cm, width=\linewidth+6cm]
\begin{figure}[H]
    \centering
    \hspace{-6cm}
    \includegraphics[width = 8.cm]{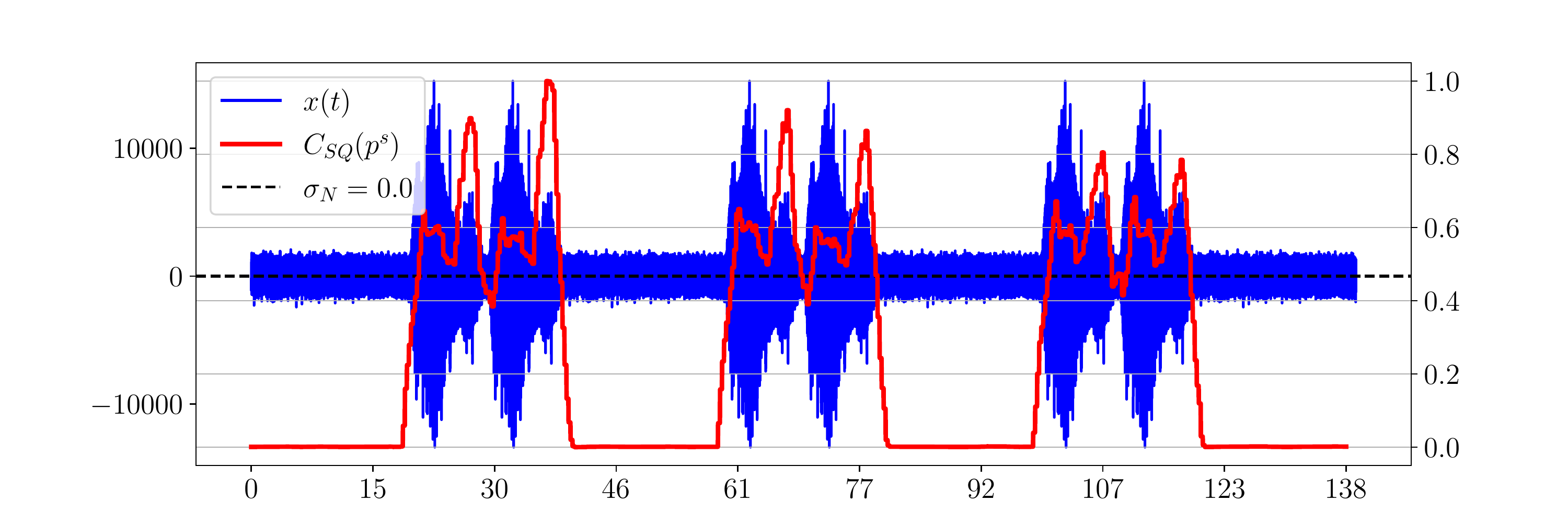}
    \hspace{-0.9cm}
    \includegraphics[width = 8.cm]{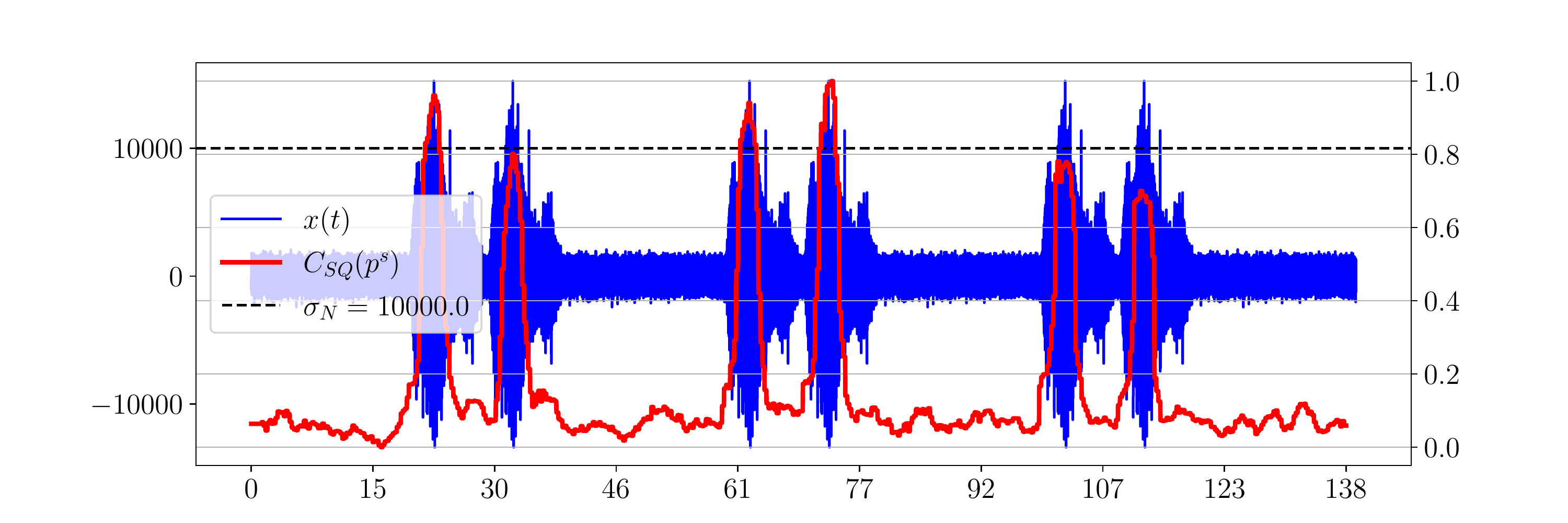}\\
    \hspace{-6cm}
    \includegraphics[width = 8.cm]{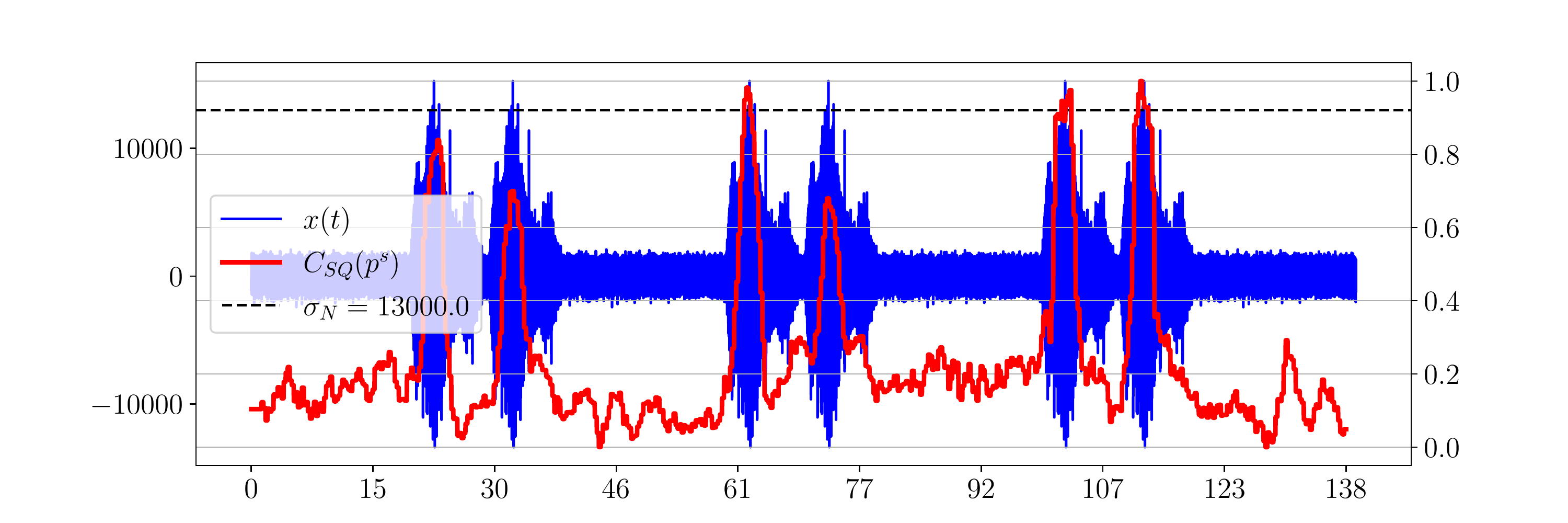}
    \hspace{-0.9cm}
     \includegraphics[width = 8cm]{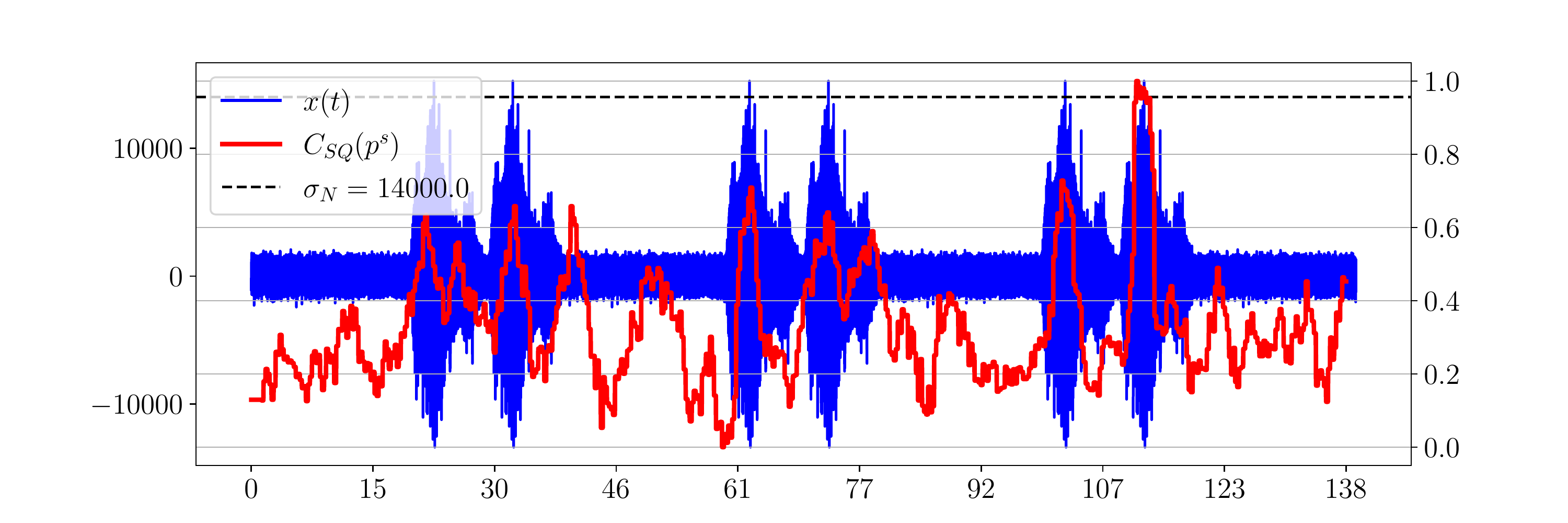}
    \caption{Graphs of statistical complexity for different levels of added noise.}
    \label{fig:sub_сomplexities_sq}
\end{figure}
\end{fullwidth}

As one can see, the statistical complexity in comparison with all other information metrics shows the best result in the problem of indicating the presence of an useful signal in the mixture, because it remains effective for small SNR, when all other characteristics no longer allow to detect a useful signal in the noisy receiving channel. 

\section{Conclusion}\label{Conclusion}

The article presents a comparison of different information criteria for indicating of the appearance of a useful signal in a heavily noisy mixture based on statistical complexity. The analytical formulas for disequilibrium and statistical complexity are obtained using entropy variation. The effectiveness of statistical complexity for two types of acoustic signals in comparison with other information metrics is shown. It occurs that the appearance of determined signal is reliably detected for a very small SNR ($\approx -15$ dB) when statistical complexity based on spectral distribution variance is used as a criterion. Both time and frequency domain are considered for entropy calculation. The criteria for signal detection in a heavy noise mixture based on time distributions are less informative than ones based on spectral distribution. The main advantage of using statistical complexity and entropy for spectral distribution as criteria lies in comparison of their calculated values with minimum (zero) and maximum (one) possible values, and also in the lack of necessity for additional selective assessment of the current noise characteristics and its evolution in time. Future work will be devoted to the research of the information criteria introduced by formulas (\ref{2DD1}) and (\ref{2DD2}) based on two- and multi-dimensional distributions and acoustic signals with more realistic background noises will be considered. 

\vspace{6pt}

The following abbreviations are used in this manuscript:\\

\noindent 
\begin{tabular}{@{}ll}
FFT & Fast Fourier Transform\\
SNR & Signal-to-noise ratio
\end{tabular}


\begin{appendices}

\appendix
\section[\appendixname~\thesection]{}
\subsection[\appendixname~\thesubsection]{}
\begin{proof}[Proof of Lemma \ref{lemma1}]

The difference between the entropies for the distributions $p_i$ and $q_i$ gives the entropy variation $\delta H$:
\begin{equation*}
\begin{array}{l}
\displaystyle\delta H = H(q+\delta q)-H(q) = -\sum_{i=1}^N (q_i+\delta q_i)\log_2(q_i+\delta q_i)+\sum_{i=1}^N q_i\log_2q_i=\\
\displaystyle=-\sum_{i=1}^N (q_i+\delta q_i)\log_2\left(q_i\left(1+\frac{\delta q_i}{q_i}\right)\right)+\sum_{i=1}^N q_i\log_2q_i.
\end{array}
\end{equation*}
The property of the logarithm of the product and the regrouping of the summands allows the chain of equations to continue as follows
\begin{equation*}
\begin{array}{l}
\displaystyle\delta H =-\sum_{i=1}^N (q_i+\delta q_i)\left(\log_2q_i+\log_2\left(1+\frac{\delta q_i}{q_i}\right)\right)+\sum_i^N q_i\log_2q_i=-\sum_{i=1}^N q_i\log_2(1+\frac{\delta q_i}{q_i})-\\
\displaystyle-\sum_{i=1}^N \delta q_i\left(\log_2q_i+\log_2\left(1+\frac{\delta q_i}{q_i}\right)\right)
\displaystyle=-\sum_{i=1}^N \delta q_i\log_2q_i-\sum_{i=1}^N (q_i+\delta q_i)\log_2\left(1+\frac{\delta q_i}{q_i}\right).
\end{array}
\end{equation*}
The first sum is equal to the difference of cross-entropy and entropy. The next transformation is decomposition into an infinite logarithm series and division of the resulting sum into two parts:
\begin{equation*}
\begin{array}{l}
\displaystyle\delta H=H(p,q)-H(q)-\sum_{i=1}^N\sum_{n=1}^\infty (q_i+\delta q_i) \frac{(-1)^{n+1}}{n\ln{2}}\left(\frac{\delta q_i}{q_i}\right)^n=H(p,q)-H(q)-\\
\displaystyle -\sum_{i=1}^N\sum_{n=1}^\infty \frac{(-1)^{n+1}\delta q_i^n}{n\ln{2}q_i^{n-1}}-\sum_{i=1}^N\sum_{n=1}^\infty \frac{(-1)^{n+1}\delta q_i^{n+1}}{n\ln{2}q_i^n}.\\
\end{array}
\end{equation*}
One summand corresponding to $n=1$ is removed from the first sum, and summation continues with $n=2$. 
\begin{equation*}
\begin{array}{l}
\displaystyle\delta H =H(p,q)-H(q)-\sum_{i=1}^N \frac{\delta q_i}{\ln 2}-\sum_{i=1}^N\sum_{n=2}^\infty \frac{(-1)^{n+1}\delta q_i^n}{n\ln{2}q_i^{n-1}}-\sum_{i=1}^N\sum_{n=1}^\infty \frac{(-1)^{n+1}\delta q_i^{n+1}}{n\ln{2}q_i^n}.\\
\end{array}
\end{equation*}
The resulting summand is zero. Shifting the summation index of the first summation results in $\delta H$ in the following form
\begin{equation*}
\begin{array}{l}
\displaystyle\delta H=H(p,q)-H(q)+\sum_{i=1}^N\sum_{n=1}^\infty \frac{(-1)^{n+1}\delta q_i^{n+1}}{(n+1)\ln{2}q_i^{n}}-\sum_{i=1}^N\sum_{n=1}^\infty \frac{(-1)^{n+1}\delta q_i^{n+1}}{n\ln{2}q_i^n}=H(p,q)-H(q)+\\
\displaystyle +\sum_{i=1}^N\sum_{n=1}^\infty \frac{(-1)^{n+1}\delta q_i^{n+1}}{\ln{2}q_i^{n}}\left(\frac{1}{n+1}-\frac{1}{n}\right)
\displaystyle =H(p,q)-H(q)-\sum_{i=1}^N\sum_{n=1}^\infty \frac{\delta q_i^{n+1}(-1)^{n+1}}{q_i^n n (n+1)\ln2}.\\

\end{array}
\end{equation*}
Another shift of the summation index leads to the equation \eqref{var_entrop}, which ends the proof of the Lemma.
\end{proof}

\end{appendices}



\bibliography{sn-bibliography.bib}


\begin{thebibliography}{25}
\ifx \bisbn   \undefined \def \bisbn  #1{ISBN #1}\fi
\ifx \binits  \undefined \def \binits#1{#1}\fi
\ifx \bauthor  \undefined \def \bauthor#1{#1}\fi
\ifx \batitle  \undefined \def \batitle#1{#1}\fi
\ifx \bjtitle  \undefined \def \bjtitle#1{#1}\fi
\ifx \bvolume  \undefined \def \bvolume#1{\textbf{#1}}\fi
\ifx \byear  \undefined \def \byear#1{#1}\fi
\ifx \bissue  \undefined \def \bissue#1{#1}\fi
\ifx \bfpage  \undefined \def \bfpage#1{#1}\fi
\ifx \blpage  \undefined \def \blpage #1{#1}\fi
\ifx \burl  \undefined \def \burl#1{\textsf{#1}}\fi
\ifx \doiurl  \undefined \def \doiurl#1{\url{https://doi.org/#1}}\fi
\ifx \betal  \undefined \def \betal{\textit{et al.}}\fi
\ifx \binstitute  \undefined \def \binstitute#1{#1}\fi
\ifx \binstitutionaled  \undefined \def \binstitutionaled#1{#1}\fi
\ifx \bctitle  \undefined \def \bctitle#1{#1}\fi
\ifx \beditor  \undefined \def \beditor#1{#1}\fi
\ifx \bpublisher  \undefined \def \bpublisher#1{#1}\fi
\ifx \bbtitle  \undefined \def \bbtitle#1{#1}\fi
\ifx \bedition  \undefined \def \bedition#1{#1}\fi
\ifx \bseriesno  \undefined \def \bseriesno#1{#1}\fi
\ifx \blocation  \undefined \def \blocation#1{#1}\fi
\ifx \bsertitle  \undefined \def \bsertitle#1{#1}\fi
\ifx \bsnm \undefined \def \bsnm#1{#1}\fi
\ifx \bsuffix \undefined \def \bsuffix#1{#1}\fi
\ifx \bparticle \undefined \def \bparticle#1{#1}\fi
\ifx \barticle \undefined \def \barticle#1{#1}\fi
\bibcommenthead
\ifx \bconfdate \undefined \def \bconfdate #1{#1}\fi
\ifx \botherref \undefined \def \botherref #1{#1}\fi
\ifx \url \undefined \def \url#1{\textsf{#1}}\fi
\ifx \bchapter \undefined \def \bchapter#1{#1}\fi
\ifx \bbook \undefined \def \bbook#1{#1}\fi
\ifx \bcomment \undefined \def \bcomment#1{#1}\fi
\ifx \oauthor \undefined \def \oauthor#1{#1}\fi
\ifx \citeauthoryear \undefined \def \citeauthoryear#1{#1}\fi
\ifx \endbibitem  \undefined \def \endbibitem {}\fi
\ifx \bconflocation  \undefined \def \bconflocation#1{#1}\fi
\ifx \arxivurl  \undefined \def \arxivurl#1{\textsf{#1}}\fi
\csname PreBibitemsHook\endcsname

\bibitem{shannon}
\begin{barticle}
\bauthor{\bsnm{Shannon}, \binits{C.E.}}:
\batitle{A mathematical theory of communication}.
\bjtitle{The Bell System Technical Journal}
\bvolume{27},
\bfpage{379}--\blpage{423}
(\byear{1948})
\end{barticle}
\endbibitem

\bibitem{ent_uni}
\begin{botherref}
\oauthor{\bsnm{Ribeiro}, \binits{M.}},
\oauthor{\bsnm{Henriques}, \binits{T.}},
\oauthor{\bsnm{Castro}, \binits{L.}},
\oauthor{\bsnm{Souto}, \binits{A.}},
\oauthor{\bsnm{Antunes}, \binits{L.}},
\oauthor{\bsnm{Costa-Santos}, \binits{C.}},
\oauthor{\bsnm{Teixeira}, \binits{A.}}:
The entropy universe.
Entropy
\textbf{23}(2)
(2021).
\doiurl{10.3390/e23020222}
\end{botherref}
\endbibitem

\bibitem{gray_entropy_2011}
\begin{bbook}
\bauthor{\bsnm{Gray}, \binits{R.M.}}:
\bbtitle{Entropy and {Information} {Theory}}.
\bpublisher{Springer},
\blocation{Boston, MA}
(\byear{2011}).
\doiurl{10.1007/978-1-4419-7970-4}.
\burl{http://link.springer.com/10.1007/978-1-4419-7970-4}
Accessed 2022-10-13
\end{bbook}
\endbibitem

\bibitem{SampleEntropy}
\begin{barticle}
\bauthor{\bsnm{Delgado-Bonal}, \binits{A.}},
\bauthor{\bsnm{Marshak}, \binits{A.}}:
\batitle{Approximate entropy and sample entropy: A comprehensive tutorial}.
\bjtitle{Entropy}
\bvolume{21},
\bfpage{541}
(\byear{2019}).
\doiurl{10.3390/e21060541}
\end{barticle}
\endbibitem

\bibitem{Shen1998RobustEE}
\begin{botherref}
\oauthor{\bsnm{Shen}, \binits{J.-L.}},
\oauthor{\bsnm{Hung}, \binits{J.-W.}},
\oauthor{\bsnm{Lee}, \binits{L.-S.}}:
Robust entropy-based endpoint detection for speech recognition in noisy
  environments.
5th International Conference on Spoken Language Processing (ICSLP 1998)
(1998)
\end{botherref}
\endbibitem

\bibitem{book_Shiryaev}
\begin{bbook}
\bauthor{\bsnm{Shiryaev}, \binits{A.N.}},
\bauthor{\bsnm{Spokoiny}, \binits{V.G.}}:
\bbtitle{Statistical Experiments and Decisions}.
\bpublisher{WORLD SCIENTIFIC}, \blocation{???}
(\byear{2000}).
\doiurl{10.1142/4247}
\end{bbook}
\endbibitem

\bibitem{Moriarty}
\begin{barticle}
\bauthor{\bsnm{Johnson}, \binits{P.}},
\bauthor{\bsnm{Moriarty}, \binits{J.}},
\bauthor{\bsnm{Peskir}, \binits{G.}}:
\batitle{Detecting changes in real-time data: a user's guide to optimal
  detection}.
\bjtitle{Philosophical Transactions of the Royal Society A: Mathematical,
  Physical and Engineering Sciences}
\bvolume{375}(\bissue{2100}),
\bfpage{16}
(\byear{2017}).
\doiurl{10.1098/rsta.2016.0298}
\end{barticle}
\endbibitem

\bibitem{mehrotra_anomaly_2017}
\begin{bbook}
\bauthor{\bsnm{Mehrotra}, \binits{K.G.}},
\bauthor{\bsnm{Mohan}, \binits{C.K.}},
\bauthor{\bsnm{Huang}, \binits{H.}}:
\bbtitle{Anomaly {Detection} {Principles} and {Algorithms}}.
\bsertitle{Terrorism, {Security}, and {Computation}}.
\bpublisher{Springer},
\blocation{Cham}
(\byear{2017}).
\doiurl{10.1007/978-3-319-67526-8}.
\burl{http://link.springer.com/10.1007/978-3-319-67526-8}
Accessed 2022-09-14
\end{bbook}
\endbibitem

\bibitem{e22080845}
\begin{botherref}
\oauthor{\bsnm{Howedi}, \binits{A.}},
\oauthor{\bsnm{Lotfi}, \binits{A.}},
\oauthor{\bsnm{Pourabdollah}, \binits{A.}}:
An entropy-based approach for anomaly detection in activities of daily living
  in the presence of a visitor.
Entropy
\textbf{22}(8)
(2020).
\doiurl{10.3390/e22080845}
\end{botherref}
\endbibitem

\bibitem{e17042367}
\begin{barticle}
\bauthor{\bsnm{Bereziński}, \binits{P.}},
\bauthor{\bsnm{Jasiul}, \binits{B.}},
\bauthor{\bsnm{Szpyrka}, \binits{M.}}:
\batitle{An entropy-based network anomaly detection method}.
\bjtitle{Entropy}
\bvolume{17}(\bissue{4}),
\bfpage{2367}--\blpage{2408}
(\byear{2015}).
\doiurl{10.3390/e17042367}
\end{barticle}
\endbibitem

\bibitem{TakuyaHorie}
\begin{barticle}
\bauthor{\bsnm{Horie}, \binits{T.}},
\bauthor{\bsnm{Burioka}, \binits{N.}},
\bauthor{\bsnm{Amisaki}, \binits{T.}},
\bauthor{\bsnm{Shimizu}, \binits{E.}}:
\batitle{Sample entropy in electrocardiogram during atrial fibrillation}.
\bjtitle{Yonago Acta Medica}
\bvolume{61}(\bissue{1}),
\bfpage{049}--\blpage{057}
(\byear{2018}).
\doiurl{10.33160/yam.2018.03.007}
\end{barticle}
\endbibitem

\bibitem{vad}
\begin{barticle}
\bauthor{\bsnm{Ramirez}, \binits{J.}},
\bauthor{\bsnm{Segura}, \binits{J.C.}},
\bauthor{\bsnm{Benitez}, \binits{C.}},
\bauthor{\bparticle{de~la} \bsnm{Torre}, \binits{A.}},
\bauthor{\bsnm{Rubio}, \binits{A.J.}}:
\batitle{A new kullback-leibler vad for speech recognition in noise}.
\bjtitle{IEEE Signal Processing Letters}
\bvolume{11}(\bissue{2}),
\bfpage{266}--\blpage{269}
(\byear{2004}).
\doiurl{10.1109/LSP.2003.821762}
\end{barticle}
\endbibitem

\bibitem{RobustEndpoint}
\begin{barticle}
\bauthor{\bsnm{Wu}, \binits{B.-F.}},
\bauthor{\bsnm{Wang}, \binits{K.-C.}}:
\batitle{Robust endpoint detection algorithm based on the adaptive
  band-partitioning spectral entropy in adverse environments}.
\bjtitle{Speech and Audio Processing, IEEE Transactions on}
\bvolume{13},
\bfpage{762}--\blpage{775}
(\byear{2005}).
\doiurl{10.1109/TSA.2005.851909}
\end{barticle}
\endbibitem

\bibitem{Weaver}
\begin{bchapter}
\bauthor{\bsnm{Weaver}, \binits{K.}},
\bauthor{\bsnm{Waheed}, \binits{K.}},
\bauthor{\bsnm{Salem}, \binits{F.}}:
\bctitle{An entropy based robust speech boundary detection algorithm for
  realistic noisy environments},
vol. \bseriesno{1},
pp. \bfpage{680}--\blpage{6851}
(\byear{2003}).
\doiurl{10.1109/IJCNN.2003.1223446}
\end{bchapter}
\endbibitem

\bibitem{lopez-ruiz_shannon_2005}
\begin{barticle}
\bauthor{\bsnm{López-Ruiz}, \binits{R.}}:
\batitle{Shannon information, {LMC} complexity and {Rényi} entropies: a
  straightforward approach}.
\bjtitle{Biophysical Chemistry}
\bvolume{115}(\bissue{2-3}),
\bfpage{215}--\blpage{218}
(\byear{2005}).
\doiurl{10.1016/j.bpc.2004.12.035}.
Accessed 2022-08-05
\end{barticle}
\endbibitem

\bibitem{catalan_features_2002}
\begin{barticle}
\bauthor{\bsnm{Catalán}, \binits{R.G.}},
\bauthor{\bsnm{Garay}, \binits{J.}},
\bauthor{\bsnm{López-Ruiz}, \binits{R.}}:
\batitle{Features of the extension of a statistical measure of complexity to
  continuous systems}.
\bjtitle{Physical Review E}
\bvolume{66}(\bissue{1}),
\bfpage{011102}
(\byear{2002}).
\doiurl{10.1103/PhysRevE.66.011102}.
Accessed 2022-08-05
\end{barticle}
\endbibitem

\bibitem{calbet_tendency_2001}
\begin{barticle}
\bauthor{\bsnm{Calbet}, \binits{X.}},
\bauthor{\bsnm{López-Ruiz}, \binits{R.}}:
\batitle{Tendency towards maximum complexity in a nonequilibrium isolated
  system}.
\bjtitle{Physical Review E}
\bvolume{63}(\bissue{6}),
\bfpage{066116}
(\byear{2001}).
\doiurl{10.1103/PhysRevE.63.066116}.
Accessed 2022-08-05
\end{barticle}
\endbibitem

\bibitem{chaos}
\begin{barticle}
\bauthor{\bsnm{Rosso}, \binits{O.}},
\bauthor{\bsnm{Larrondo}, \binits{H.}},
\bauthor{\bsnm{Martin}, \binits{M.T.}},
\bauthor{\bsnm{Plastino}, \binits{A.}},
\bauthor{\bsnm{Fuentes}, \binits{M.}}:
\batitle{Distinguishing noise from chaos}.
\bjtitle{Physical review letters}
\bvolume{99},
\bfpage{154102}
(\byear{2007}).
\doiurl{10.1103/PhysRevLett.99.154102}
\end{barticle}
\endbibitem

\bibitem{LAMBERTI2004119}
\begin{barticle}
\bauthor{\bsnm{Lamberti}, \binits{P.W.}},
\bauthor{\bsnm{Martin}, \binits{M.T.}},
\bauthor{\bsnm{Plastino}, \binits{A.}},
\bauthor{\bsnm{Rosso}, \binits{O.A.}}:
\batitle{Intensive entropic non-triviality measure}.
\bjtitle{Physica A: Statistical Mechanics and its Applications}
\bvolume{334}(\bissue{1}),
\bfpage{119}--\blpage{131}
(\byear{2004}).
\doiurl{10.1016/j.physa.2003.11.005}
\end{barticle}
\endbibitem

\bibitem{rosso}
\begin{barticle}
\bauthor{\bsnm{Zunino}, \binits{L.}},
\bauthor{\bsnm{Soriano}, \binits{M.C.}},
\bauthor{\bsnm{Rosso}, \binits{O.A.}}:
\batitle{Distinguishing chaotic and stochastic dynamics from time series by
  using a multiscale symbolic approach}.
\bjtitle{Phys. Rev. E}
\bvolume{86},
\bfpage{046210}
(\byear{2012}).
\doiurl{10.1103/PhysRevE.86.046210}
\end{barticle}
\endbibitem

\bibitem{features}
\begin{botherref}
\oauthor{\bsnm{Li}, \binits{Z.}},
\oauthor{\bsnm{Li}, \binits{Y.}},
\oauthor{\bsnm{Zhang}, \binits{K.}}:
A feature extraction method of ship-radiated noise based on fluctuation-based
  dispersion entropy and intrinsic time-scale decomposition.
Entropy
\textbf{21}(7)
(2019).
\doiurl{10.3390/e21070693}
\end{botherref}
\endbibitem

\bibitem{plane}
\begin{barticle}
\bauthor{\bsnm{Dai}, \binits{Y.}},
\bauthor{\bsnm{Zhang}, \binits{H.}},
\bauthor{\bsnm{Mao}, \binits{X.}},
\bauthor{\bsnm{Shang}, \binits{P.}}:
\batitle{Complexity–entropy causality plane based on power spectral entropy
  for complex time series}.
\bjtitle{Physica A: Statistical Mechanics and its Applications}
\bvolume{509},
\bfpage{501}--\blpage{514}
(\byear{2018}).
\doiurl{10.1016/j.physa.2018.06.081}
\end{barticle}
\endbibitem

\bibitem{US_patent}
\begin{botherref}
\oauthor{\bsnm{Quazi}, \binits{A.H.}}:
Method for detecting acoustic signals from an underwater source.
Google Patents
(1997).
\url{https://patents.google.com/patent/US5668778A/en}
\end{botherref}
\endbibitem

\bibitem{SID}
\begin{barticle}
\bauthor{\bsnm{{Chein-I Chang}}}:
\batitle{An information-theoretic approach to spectral variability, similarity,
  and discrimination for hyperspectral image analysis}.
\bjtitle{IEEE Transactions on Information Theory}
\bvolume{46}(\bissue{5}),
\bfpage{1927}--\blpage{1932}
(\byear{2000}).
\doiurl{10.1109/18.857802}.
Accessed 2022-10-20
\end{barticle}
\endbibitem

\bibitem{fdivergence}
\begin{botherref}
\oauthor{\bsnm{Sason}, \binits{I.}}:
On f-divergences: Integral representations, local behavior, and inequalities.
Entropy
\textbf{20}(5)
(2018).
\doiurl{10.3390/e20050383}
\end{botherref}
\endbibitem

\end{thebibliography}


%


\end{document}